\documentclass[onecolumn,showpacs,amsmath,amssymb,noshowpacs]
{revtex4-2}

\usepackage{array} % for better arrays (eg matrices) in maths
\usepackage{braket}

\usepackage{tikz}
\usetikzlibrary{fadings}
\usetikzlibrary{patterns}
\usetikzlibrary{shadows.blur}
\usetikzlibrary{shapes}

\usepackage{definitions}%

\newcommand{\zz}[1]{{#1}}
\newcommand{\kett}[1]{| #1 \rangle\rangle}
\newcommand{\bbra}[1]{\langle \langle #1 |}
\newcommand{\bbrakett}[2]{\langle \langle #1 | #2 \rangle \rangle}
\begin{document}

\title{Hall conductivity as the topological invariant in magnetic Brillouin zone}% Force line breaks with \\

\author{M. Suleymanov}
\affiliation{Physics Department, Ariel University, Ariel 40700, Israel}

\author{M. A. Zubkov}
\affiliation{Physics Department, Ariel University, Ariel 40700, Israel}

\author{C. X. Zhang}
\affiliation{Wuerzburg University, Am Hubland
	97074 Wuerzburg, Germany }

\begin{abstract}
Hall conductivity for the intrinsic quantum Hall effect in homogeneous systems is given by the topological invariant composed of the Green function depending on momentum of quasiparticle. This expression reveals correspondence with the mathematical notion of the degree of mapping. A more involved situation takes place for the Hall effect in the presence of external magnetic field. In this case the mentioned expression remains valid if the Green function is replaced by its Wigner transformation while ordinary products are replaced by the Moyal products. Such an expression, unfortunately, is much more complicated and might be useless for the practical calculations. Here we represent the alternative representation for the Hall conductivity of a uniform system in the presence of constant magnetic field. The Hall conductivity is expressed through the Green function taken in Harper representation, when its nonhomogeneity is attributed to the matrix structure while functional dependence is on one momentum that belongs to magnetic Brillouin zone. Our results were obtained for the non - interacting systems. But we expect that they remain valid for the interacting systems as well. We, therefore, propose the hypothesis that the obtained expression may be used for the topological description of fractional quantum Hall effect.
\end{abstract}

\pacs{}% PACS, the Physics and Astronomy

%\tableofcontents

%\tableofcontents
%\newpage
\maketitle
\tableofcontents

\section{Introduction}

%{The}
The Quantum Hall effect (QHE) is widely believed to be related to concepts of topology and geometry \cite{nagaosa}.
The first topological expression for the QHE conductivity has been proposed in \cite{TKNN} for the ideal two dimensional non-interacting condensed matter systems in the presence of constant external magnetic field. This expression is proportional to the TKNN invariant, which is the integral of Berry curvature in magnetic Brillouin zone
over the occupied electronic states \cite{Fradkin,Tong:2016kpv,Hatsugai}. This expression also remains valid for the description of integer intrinsic QHE in 2d topological insulators. Besides, this approach has been extended to the three space-dimensional (3d) topological insulators
(see, for example, \cite{Hall3DTI}). The introduction
of weak interactions as well as disorder does not affect total Hall conductivity. It is important, therefore, to express Hall conductivity
through the Green functions that are well - defined within interacting theory.
Such an expression for the conductivity of intrinsic anomalous QHE (AQHE) in $2+1$ D systems
has been given through the Green functions in \cite{Matsuyama:1986us,Imai:1990zz,Volovik0} (see also Chapter 21.2.1 in \cite{Volovik2003}).
The extension of such a construction to various 3d systems has also been proposed \cite{Z2016_1}. The resulting expression allows to describe the AQHE in Weyl semimetals \cite{semimetal_effects10,semimetal_effects11,semimetal_effects12,semimetal_effects13,Zyuzin:2012tv,tewary}. The similar topological invariants have also been discussed
in \cite{Gurarie2011,EssinGurarie2011}. The AQHE conductivity is given by the expressions of  \cite{Matsuyama:1986us,Volovik0,Volovik2003,Z2016_1} through the two - point Green functions also for the interacting systems. The general proof of this statement has been given in \cite{ZZ2019_0}.

In \cite{ZW2019} the construction of \cite{Matsuyama:1986us,Imai:1990zz,Volovik0} was extended to the essentially non - homogeneous systems. It appears that the Hall conductivity is expressed through the Wigner transformed two - point Green function.
This gives an alternative proof that disorder does not affect the total QHE conductivity (although the local Hall current is pushed by disorder towards the boundary of the sample). It is worth mentioning that the role of disorder
in QHE has been widely discussed in the past \cite{Fradkin,TKNN2,QHEr,Tong:2016kpv,Zheng+Ando2002}. The absence of corrections due to weak Coulomb interactions to the QHE ferromagnetic metal was discussed in \cite{nocorrectionstoQHE}. Inter - electron interactions and their relation to QHE have also been discussed long time ago (see, for example,  \cite{Altshuler0,Altshuler,Ishikawa:2003tu,Imai:1990zz}). The proof of the absence of radiative corrections to Hall conductivity (to all orders in perturbation theory) in the presence of external magnetic field has been given in \cite{ZZ2019_1,ZZ2021} (see also \cite{parity_anomaly}).

Unfortunately, expression proposed in \cite{ZW2019} does not look useful for practical calculations since it contains the Moyal product of functions defined in phase space.
Another representation of the QHE conductivity through the two - point Green functions has been given in \cite{Imai:1990zz} (see also references therein). In principle, the idea of \cite{Imai:1990zz} is somehow similar to that of \cite{ZW2019} and of the present paper: the final form of topological expression for the Hall conductivity resembles the expression for the degree of mapping, where the mapping is defined by the Green function. In \cite{ZW2019} this expression is modified replacing the ordinary product by Moyal product, while the Green function is replaced by its Wigner transform. In \cite{Imai:1990zz} the ordinary matrix product is used. However, the Green function is taken in the specific representation, when it is represented by the infinite dimension matrix depending on momentum. Otherwise the topological invariant of \cite{Imai:1990zz} is similar to Eq. (\ref{NT}) of the present paper. Since the matrix is infinite dimensional, such an expression looks useless for the computational purposes. The advantage of  expression proposed by us in the present paper is that it is composed of the $N\times N$ matrices with finite $N$. Those matrices depend on momentum.

Although we derive our expression for the Hall conductivity for the non - interacting model, we suppose that it remains
valid also in the presence of interactions, when the non - interacting Green function is replaced by the complete two point Green function with the interaction corrections.
It is well - known that interactions are able to lead the fermionic system in the presence of external magnetic field to the fractional QHE phases \cite{Tong:2016kpv}. We propose the hypothesis that our expression might be
used for the topological description of the QHE in these phases. The fractional QHE (FQHE) may be observed when disorder is decreased. Then the additional plateaux emerge in the quantum Hall conductivity. This effect was discovered first for the fractional conductivity equal to $\nu=1/3$ of the Klitzing constant $e^2/h$ \cite{PhysRevLett.48.1559}.  It was supposed by Laughlin that the origin of the observed FQHE with $\nu = 1/3$, as well as any $n = 1/q$ with odd integer $q$ is due to the formation of the correlated incompressible electron liquid with exotic properties \cite{PhysRevLett.50.1395}. Later the theoretical description of the other types of FQHE was given including the FQHE with $\nu=2/5$ and $\nu=3/7$, as a part of the $p/(2sp\pm 1)$ series (  $s,p\in\Z$). It is widely believed that the FQHE  may be explained by the so-called {\itshape composite-fermion} theory, in which the FQHE is viewed as an integer QHE of a novel quasi-particle that consists of an electron that “captures” an even number of magnetic flux quanta \cite{PhysRevLett.63.199},\cite{PhysRevB.41.7653}.

Notice that the AQHE existing without magnetic field can also exist in fractional form \cite{FAQHE}. The effects of interactions within the 2d topological insulators were considered
in \cite{2DTI_corr}. In graphene - like sytems relation of Coulomb interactions to the renormalization of Fermi velocity was studied, for example, in \cite{Tang2018_Science}.
Various questions related to interaction effects in 2D systems have been discussed in  \cite{corr_2d,corr_2d_2,AQHE_no_corr}.
The interaction corrections in 3D Weyl semimetals are discussed in
  \cite{corr_WSM1,corr_WSM2}.
In the present paper we consider the tight - binding models in 2d. Similar tight - binding models have been widely considered in the past (see, for example, in \cite{Z2016_1,tb2d,tb2d2,tb2d3,corr_2d} and references therein). For the description of various 3d tight - binding models see \cite{tb1,tb2,tb3,tb4,tb5}.

The paper is organized as follows.
In Sect. \ref{SectionII}  we remind the reader the basic notions of the field theoretical description of condensed matter systems. Here we refer to Appendixes \ref{AppendixA},\ref{AppendixB},\ref{AppendixC} for certain technical details that are included to the present paper for completeness. In Sect. \ref{SectionIII} we consider the simplest 2d tight - binding model defined on rectangular lattice in the presence of constant external magnetic field. By consideration of this model we illustrate our general derivation for the topological expression of Hall conductivity. This derivation remains valid for the wide ranges of tight - binding models defined on rectangular lattices. Certain results related to this section are placed in Appendix \ref{AppendixD} and Appendix \ref{AppendixE}.
In Sect. \ref{SectionIV} we illustrate the  obtained results by numerical calculation of Hall conductivity for the two particular cases. We demonstrate that the derived topological expression for the Hall conductivity reproduces the known result obtained using the Diophantine equation.
In Sect. \ref{SectionV} we end with the conclusions and discuss the possible outcome of the present research.

\section{Basic notions of path integral formulation for the condensed matter systems}
\label{SectionII}

\subsection{Functional path integral - Partition function}

In this section we remind briefly the basic notions of the field theoretic formulation for the condensed matter systems. For completeness we accumulate technical details related to the coherent states in Appendix \ref{AppendixA} (bosons) and Appendix \ref{AppendixB} (fermions). The Gaussian integrals are considered briefly in Appendix \ref{AppendixC}.

We start from the Hibbs distribution for the field system with Hamiltonian $\^H$:
\be
\^\rho= \frac{1}{\cZ} e^{-\beta\^H}
\ee
The path integral representation of partition function of a fermionic system is obtained using the formalism of coherent states described in details in Appendix \ref{AppendixB}. For the notations of the coherent states and their relation to the creation and annihilation operators see also the mentioned Appendix.
Adopting a shorthand notation $\psi^{(n)}=\{\psi^{(n)}_i\}$, and
$
\bar\psi^{(n)}\psi^{(m)}\equiv \sum_i \bar\psi^{(n)}_i\psi^{(m)}_i
$
we obtain the following representation for the partition function
\be
\cZ =& \tr e^{-\beta\^H}=
\sum_n \bra{n} e^{-\beta\^H} \ket{n}\\
=& \int d\(\bar\psi,\psi\)
e^{-\bar\psi\psi}
\bra{-\psi} e^{-\beta\^H} \ket{\psi}\\
=& \int d\(\bar\psi^{(0)},\psi^{(0)}\)  e^{-\bar\psi^{(0)}\psi^{(0)}}
\int d\(\bar\psi^{(M-1)},\psi^{(M-1)}\)  e^{-\bar\psi^{(M-1)}\psi^{(M-1)}}...
\int d\(\bar\psi^{(1)},\psi^{(1)}\)  e^{-\bar\psi^{(1)}\psi^{(1)}}
\\
& \bra{-\psi^{(0)}}
e^{-\frac{\beta}{M}\^H}
\ket{\psi^{(M-1)}}
\bra{\psi^{(M-1)}}
e^{-\frac{\beta}{M}\^H}
\ket{\psi^{(M-2)}}...
\bra{\psi^{(1)}}
e^{-\frac{\beta}{M}\^H}
\ket{\psi^{(0)}}\\
\ee
In the limit
\be
M\ra \infty \tab \ep=\frac{\beta}{M} \ra 0
\ee
for a normally ordered Hamiltonian $\^H\(\^a^\+,\^a\)$:
\be
\bra{\psi^{(n+1)}} e^{-\ep H(a^\+,a)} \ket{\psi^{(n)}}=
%\bra{\psi^{(n+1)}} \(1-\ep H(a^\+,a)\) \ket{\psi^{(n)}}=
e^{\bar\psi^{(n+1)}\psi^{(n)}} e^{-\ep H\(\bar\psi^{(n+1)},\psi^{(n)}\)}
\ee
and
\be
\cZ &= \prod_{n=0}^{M-1} \int_{\psi^{(M)}=-\psi^{(0)}} d\(\bar\psi^{(n)},\psi^{(n)}\)
e^{\bar\psi^{(n)}\psi^{(n+1)}-\bar\psi^{(n)}\psi^{(n)}-
	\ep H\(\bar\psi^{(n)},\psi^{(n+1)} \)}\\
&=  \int_{\psi^{(M)}=-\psi^{(0)}}
\(\prod_{n=0}^{M-1} d\(\bar\psi^{(n)},\psi^{(n)}\) \)
e^{\sum_{n=0}^{M-1}
\[	\bar\psi^{(n+1)}\psi^{(n)}-\bar\psi^{(n)}\psi^{(n)}-
	\ep H\(\bar\psi^{(n)},\psi^{(n+1)} \)\]}
\ee

\be
D\(\bar\psi,\psi\)\equiv \prod_{n=0}^{M-1} d\(\bar\psi^{(n)},\psi^{(n)}\)
\ee

\be
\cZ
&=  \int_{\psi^{(M)}=-\psi^{(0)}}
 D\(\bar\psi,\psi\)
e^{-\sum_{n=0}^{M-1}
	\ep\[	-\frac{\bar\psi^{(n+1)}-\bar\psi^{(n)}}{\ep}\psi^{(n)}+
	 H\(\bar\psi^{(n)},\psi^{(n+1)} \)\]} \\
&=\int_{\psi{(\beta)}=-\psi{(0)}}
D\(\bar\psi,\psi\)
e^{-\int\limits_{0}^{\beta}
d\tau\[	\bar\psi(\tau)\pd_\tau\psi(\tau)+
	H\(\bar\psi(\tau),\psi(\tau) \)
\]}
\ee
The boundary conditions here have the form
\be
\psi(\beta)=-\psi(0) \tab \bar\psi(\beta)=-\bar\psi(0)
\ee
We rewrite the above expression as
\be
\cZ
&=\int D(\bar\psi,\psi) e^{-S[\bar\psi,\psi]}
\ee
Adding chemical potential, we obtain
\be
S[\bar\psi,\psi]=
\int_0^\beta d\tau \Big(\bar\psi(\tau)\pd_\tau\psi(\tau)+H(\bar\psi(\tau),\psi(\tau))-\mu N(\bar\psi(\tau),\psi(\tau))\Big)
\ee
with
\be
H(\bar\psi(\tau),\psi(\tau))=\frac{\bra{\psi(\tau)}\^H(a^\dag,a)\ket{\psi(\tau)}}{\braket{\psi(\tau)|\psi(\tau)}} \tab
N(\bar\psi(\tau),\psi(\tau))=\frac{\bra{\psi(\tau)}\^N(a^\dag,a)\ket{\psi(\tau)}}{\braket{\psi(\tau)|\psi(\tau)}}
\ee
We can always choose to work with normalized coherent states (see Appendix \ref{AppendixN} for definition) with $\bbrakett{\psi}{\psi}=1$:
\be
H(\bar\psi(\tau),\psi(\tau))=
\bbra{\psi(\tau)}\^H(a^\dag,a)\kett{\psi(\tau)}
\tab
N(\bar\psi(\tau),\psi(\tau))=
\bbra{\psi(\tau)}\^N(a^\dag,a)\kett{\psi(\tau)}
\ee
Without interactions we have
\be
\^H(a^\dag,a)=\sum_{ij} h_{ij}\^a^\+_i\^a_j
\ee
Particle number operator
\be
N(a^\+,a)=\sum_i a^\+_i a_i
\ee
in space representation receives the form
\be
N(a^\+,a)=\sum_x a^\+_x a_x
\ee
However, for
\be
\bar\psi(\tau)\pd_\tau\psi(\tau)=\sum_i \bar\psi_i(\tau)\pd_\tau\psi_i(\tau)
\ee
we have
\be
& \bra{\psi(\tau)} \pd_\tau \ket{\psi(\tau)}=\\
&=\bra{\psi(\tau)}\pd_\tau \exp\(-\sum_i\psi_i(\tau)a^\+_i\)\ket{0}=
\bra{\psi(\tau)} \(-\sum_i\pd_\tau \psi_i(\tau)a^\+_i\)
 \exp\(-\sum_i\psi_i(\tau)a^\+_i\)\ket{0}=\\
 &=\bra{\psi(\tau)} \(\sum_i a^\+_i \pd_\tau \psi_i(\tau)\)\ket{\psi(\tau)}
% =\bra{\psi(\tau)}\(\sum_i a^\+_i \pd_\tau a_i\)\ket{\psi(\tau)}
 =\bra{\psi(\tau)}\(\sum_i \bar\psi_i(\tau) \pd_\tau \psi_i(\tau)\)\ket{\psi(\tau)} =\\
&= \sum_i \Big( \bar\psi_i(\tau) \pd_\tau \psi_i(\tau) \Big) \braket{\psi(\tau)|\psi(\tau)}=
\Big( \bar\psi(\tau)\pd_\tau\psi(\tau) \Big)
\braket{\psi(\tau)|\psi(\tau)}
\ee
i.e.
\be
& \bra{\psi(\tau)} \pd_\tau \ket{\psi(\tau)}=
 \bar\psi(\tau)\pd_\tau\psi(\tau) \braket{\psi(\tau)|\psi(\tau)}
\ee
\zz{Using the normalized coherent states we would receive an extra term originated from the derivative of $\exp\(-\frac{1}{2}\sum_i \bar\psi_i\psi_i\)$
and we obtain
\be
& \bbra{\psi(\tau)} \pd_\tau \kett{\psi(\tau)}=
\bar\psi(\tau)\pd_\tau\psi(\tau)-\frac{1}{2}\pd_\tau (\bar\psi(\tau)\psi(\tau))
\ee
However, in view of the anti - periodic boundary conditions the integral over $\tau$ of the second term in this expression disappears.}
Therefore, we represent
\be
 S[\bar\psi,\psi]=\int_0^\beta d\tau
\bbra{\psi(\tau)}\(\pd_\tau+
\^H(a^\dag,a)-\mu \^N(a^\dag,a)
\)\kett{\psi(\tau)}
\label{th_avg}
\ee
We may define
\be
\^Q(a^\+,a)=\pd_\tau+
\^H(a^\dag,a)-\mu \^N(a^\dag,a)
\ee
and represent
\be
S[\bar\psi,\psi]=\int_0^\beta d\tau
\bbra{\psi(\tau)}
\^Q\kett{\psi(\tau)}=
\int_0^\beta d\tau
\Tr\(\^Q \kett{\psi(\tau)}\bbra{\psi}(\tau)\)
\label{S_Q}
\ee

%\be \bra{\psi}\^Q \ket{\psi}= \sum_{ij} \braket{\psi|i}\bra{i}\^Q \ket{j}\braket{j|\psi}= \sum_{ij}  \bar\psi_i Q_{ij} \psi_j= \bm{\bar\psi}^T \bm{Q} \bm{\psi} \ee
%\be  \bra{\psi}\^Q \ket{\psi}= \sum_{i} \bra{i}\^Q \ket{\psi}\braket{\psi|i}= \Tr\(\^Q \ket{\psi}\bra{\psi}\) \ee

%\subsection{Matsubara "frequencies"}
\subsection{Matsubara "frequencies"}
The Fourier transform in  $\tau$, which may be useful, is defined as follows
\be
\psi(\tau)=\sum_{\om_n} \psi(\om_n) e^{-i\om_n \tau}, \tab\tab
\psi(\om_n)=\frac{1}{\beta} \int_0^\beta d\tau \psi(\tau) e^{+i\om_n\tau}\\
\bar\psi(\tau)=\sum_{\om_n} \bar\psi(\om_n) e^{+i\om_n \tau}, \tab\tab
\bar\psi(\om_n)=\frac{1}{\beta} \int_0^\beta d\tau \psi(\tau) e^{-i\om_n\tau}
\ee
For bosons $\om_n=2n\pi T$, for fermions $\om_n=(2n+1)\pi T$\\ \\
Recalling that
\be
\bar\psi(\tau) \psi(\tau)=
\sum_i \bar\psi_i(\tau) \psi_i(\tau)=
\sum_x \bar\psi_x(\tau) \psi_x(\tau)
\ee
and, since
$
\int_0^\beta d\tau e^{i\(\om_n-\om_m\)\tau}=\beta\de_{mn}
$,
we may write
\be
&\int_0^\beta d\tau\bbra{\psi(\tau)}
\sum_i a^\+_i a_i
 \kett{\psi(\tau)}=
\int_0^\beta d\tau \bar\psi(\tau) \psi(\tau)=\\
&=\int_0^\beta d\tau
\sum_{\om_n} \bar\psi(\om_n) e^{+i\om_n \tau}
\sum_{\om_m} \psi(\om_m) e^{-i\om_m \tau}=
\beta\sum_{\om_n} \bar\psi(\om_n) \psi(\om_n)=\\
&= \beta \sum_{\om_n}\bbra{\psi(\om_n)}
\sum_i a^\+_i a_i
\kett{\psi(\om_n)}
\ee
and
\be
&\int_0^\beta d\tau \bar\psi(\tau) \pd_\tau \psi(\tau)=
\int_0^\beta d\tau \sum_i \bar\psi_i(\tau) \pd_\tau \psi_i(\tau)=
\beta\sum_{\om_n} \(-i\om_n\)\sum_i \bar\psi_i(\om_n) \psi_i(\om_n)=\\
&=\beta\sum_{\om_n}
\bbra{\psi(\om_n)}
\(-i\om_n\) \sum_i a^\+_i a_i
\kett{\psi(\om_n)}
\ee
Hence, the action
\be
S[\bar\psi,\psi]=
\int_0^\beta d\tau \Big(\bar\psi(\tau)\pd_\tau\psi(\tau)+H(\bar\psi(\tau),\psi(\tau))-\mu N(\bar\psi(\tau),\psi(\tau))\Big)
\ee
may be written as
\be
S[\bar\psi,\psi]=&
\beta\sum_{\om_n}
\Big(\bar\psi(\om_n)\(-i\om_n\)\psi(\om_n) +
H(\bar\psi(\om_n),\psi(\om_n))-\mu N(\bar\psi(\om_n),\psi(\om_n))\Big)=\\
&\beta\sum_{\om_n} \bbra{\psi(\om_n)}\(\(-i\om_n  -\mu\) \^N(a^\dag,a)+
\^H(a^\dag,a)
\)\kett{\psi(\om_n)}=\\
&=\beta\sum_{\om_n}
\bbra{\psi(\om_n)}\^Q(\om_n)\kett{\psi(\om_n)}=
\beta\sum_{\om_n}\Tr\(\^Q(\om_n)\kett{\psi(\om_n)}\bbra{\psi(\om_n)}\)
\ee
with
\be
\^Q_{(\om_n)}\(a^\+,a\)\equiv \(-i\om_n  -\mu\) \^N(a^\dag,a) +
\^H(a^\+,a)=\sum_{ij} a^\+_i Q_{ij}\(\om_n\) a_j
\ee
Without interactions,
$\^H(a^\dag,a)=\sum_{ij} h_{ij}\^a^\+_i\^a_j$
\be
\^Q_{(\om_n)}\(a^\+,a\)=\sum_{ij} a^\+_i Q_{ij}\(\om_n\) a_j
\ee
where
\be
Q(\om_n)_{ij}=\(-i\om_n  -\mu\)\de_{ij}+ h_{ij}
\ee
and we have
\be
S[\bar\psi,\psi]=
\beta\sum_{\om_m}\sum_{ij} \bar\psi_i(\om_m) Q_{ij}(\om_m) \psi_j(\om_m)=
\beta\sum_{\om_m} \bar{\bm\psi}^T(\om_m) \bm Q(\om_m) \bm{\psi}(\om_m)
\ee

\subsection{Thermal average}
Thermal average of operator $O$ is defined as
\be
\braket{\^O} =& \tr \(\^O\^\rho\)=&\frac{1}{\cZ} \tr \(\^O e^{-\beta\^H}\)
\ee
we represent
\be
\tr \(\^O e^{-\beta\^H}\)=
\tr \(e^{-\frac{n\beta}{M}\^H}\^O e^{-\frac{(M-n)\beta}{M}\^H}\) \tab n\in 0,..,M
\ee
and obtain
\be
\braket{\^O} =& \tr \(\^O\^\rho\)=\\
=&\frac{1}{\cZ} \tr \(\^O e^{-\beta\^H}\)=
\sum_n \bra{n} \^O e^{-\beta\^H} \ket{n}\\
=& \frac{1}{\cZ}\int d\(\bar\psi,\psi\)
e^{-\bar\psi\psi}
\bra{-\psi} \^O e^{-\beta\^H} \ket{\psi}\\
=& \frac{1}{\cZ}\int d\(\bar\psi^{(0)},\psi^{(0)}\)  e^{-\bar\psi^{(0)}\psi^{(0)}}
\int d\(\bar\psi^{(M-1)},\psi^{(M-1)}\)  e^{-\bar\psi^{(M-1)}\psi^{(M-1)}}...
\int d\(\bar\psi^{(1)},\psi^{(1)}\)  e^{-\bar\psi^{(1)}\psi^{(1)}}
\\
& \bra{-\psi^{(0)}}
\^O\(a^\+, a \)
% e^{-\frac{\beta}{M}\^H\(a^\+ a \)}
\ket{\psi^{(M-1)}}
\bra{\psi^{(M-1)}}
e^{-\frac{\beta}{M}\^H\(a^\+ a \)}
\ket{\psi^{(M-2)}}...
\bra{\psi^{(1)}}
e^{-\frac{\beta}{M}\^H\(a^\+ a \)}
\ket{\psi^{(0)}}\\
\ee
Here $O$ is a function of operators $a^+, a$ ordered normally.
Due to trace properties
\be
\braket{\^O} =& \tr \(\^O\^\rho\)=\\
=& \frac{1}{\cZ}\int d\(\bar\psi^{(0)},\psi^{(0)}\)  e^{-\bar\psi^{(0)}\psi^{(0)}}
\int d\(\bar\psi^{(M-1)},\psi^{(M-1)}\)  e^{-\bar\psi^{(M-1)}\psi^{(M-1)}}...
\int d\(\bar\psi^{(1)},\psi^{(1)}\)  e^{-\bar\psi^{(1)}\psi^{(1)}}
\\
& \bra{-\psi^{(0)}}
e^{-\frac{\beta}{M}\^H}
\ket{\psi^{(M-1)}}
\bra{\psi^{(M-1)}}
e^{-\frac{\beta}{M}\^H}
\ket{\psi^{(M-2)}}...
\bra{\psi^{(n+1)}} \^O \ket{\psi^{(n)}}
...
\bra{\psi^{(1)}}
e^{-\frac{\beta}{M}\^H}
\ket{\psi^{(0)}}\\
\ee
Here
\be
\bra{\psi^{(n+1)}} \^O\(a^\+, a\) \ket{\psi^{(n)}}=
\bra{\psi^{(n+1)}} O\(\bar\psi^{(n+1)}, \psi^{(n)}\) \ket{\psi^{(n)}}=
O\(\bar\psi^{(n+1)} , \psi^{(n)}\) \braket{\psi^{(n+1)} | \psi^{(n)}}
\ee
In the last step we assumed that $O\(\bar\psi^{(n+1)}, \psi^{(n)}\)$ contains only a balanced number of $\bar\psi$ and $\psi$, hence, the sign remains unchanged

\be
\braket{\^O}
&=\frac{1}{\cZ}\int D(\bar\psi,\psi) O\(\bar\psi(\tau),\psi(\tau) \) e^{-\int_0^\beta d\tau \Big(\bar\psi(\tau)\pd_\tau\psi(\tau)+H(\bar\psi(\tau),\psi(\tau))-\mu N(\bar\psi(\tau),\psi(\tau))\Big)}
\ee
where $\tau \in \(0,\beta\) $\\
Hence, we may write
\be
\braket{\^O}
&=
\frac{1}{\cZ}
\int D(\bar\psi,\psi)
\Bigg[ \frac{1}{M}\sum_i^M O\(\bar\psi(\tau_i),\psi(\tau_i) \) \Bigg]
e^{-\int_0^\beta d\tau \Big(\bar\psi(\tau)\pd_\tau\psi(\tau)+H(\bar\psi(\tau),\psi(\tau))-\mu N(\bar\psi(\tau),\psi(\tau))\Big)}=\\
&= \frac{1}{\cZ}
\int D(\bar\psi,\psi)
\Bigg[\frac{1}{\beta}\int_0^\beta d\tau O\(\bar\psi(\tau),\psi(\tau) \) \Bigg]
e^{-\int_0^\beta d\tau \Big(\bar\psi(\tau)\pd_\tau\psi(\tau)+H(\bar\psi(\tau),\psi(\tau))-\mu N(\bar\psi(\tau),\psi(\tau))\Big)}
\ee
that is
\be
\braket{\^O}
&= \frac{1}{\cZ}
\int D(\bar\psi,\psi)
\Bigg[\frac{1}{\beta}\int_0^\beta d\tau O\(\bar\psi(\tau),\psi(\tau) \) \Bigg]
e^{-S[\bar\psi,\psi]}
\ee

\subsection{Green's function}
\subsubsection{$\tau$ representation}
In case of the Hamiltonian with the following form
\be
\^H(a^\+,a)=\sum_{ij} a_i^\+ h_{ij} a_j
\ee
\zz{let us denote
\be \^Q=\sum_i a_i^+ \partial_\tau a_i +\^H(a^\+,a) =  \sum_{ij} a_i^\+ {\bf Q}_{ij} a_j\ee
with
$$
{\bf Q}_{ij} = \delta_{ij} \partial_\tau + h_{ij}
$$
the action may be written as follows
\begin{eqnarray}
S[\bar\psi,\psi]&=&=\int_0^\beta d\tau
\bbra{\psi(\tau)}
\^Q\kett{\psi(\tau)}=
\int_0^\beta d\tau
\Tr\(\^Q \kett{\psi(\tau)}\bbra{\psi}(\tau)\) \nonumber\\& = &
\int_0^\beta d\tau
\bm{\bar\psi}^T(\tau)\bm{Q} \bm{\psi}(\tau)=
\int_0^\beta d\tau
\sum_{xy}\bar\psi(\tau)_x Q_{xy} \psi(\tau)_y\nonumber\\ & = &
\int_0^\beta d\tau
\sum_{xy}\bra{\psi(\tau)} x\rangle {\bf Q}_{xy} \langle y \ket{ \psi(\tau)}=\int_0^\beta d\tau
\Tr\(\^{\bf Q} \ket{\psi(\tau)}\bra{\psi}(\tau)\)
\end{eqnarray}
Here
$$
\^{\bf Q} = \sum_{xy}\ket{x}  {\bf Q}_{xy} \bra{y}
$$
Notice, that $\bf Q$ represents only the first term in expansion of $Q$:
$$
\hat{Q} = \sum_{xy} a_x^\+ {\bf Q}_{xy} a_y =  \sum_{xy}{\bf Q}_{xy} a_x^\+ (\ket{0}\bra{0} + \sum_z \ket{z}\bra{z} + \frac{1}{2!} \sum_{zw}\ket{zw}\bra{zw}+...) a_y
$$
Namely, $\hat{\bf Q}$ is a restriction of $\hat{Q}$ to the subspace of Fock space. This subspace is spanned on the one - particle states.}
{Generating function for the Green functions (i.e. partition function with sources) has the form
\begin{eqnarray}
\cZ[\bar\eta,\eta]
&=\int D(\bar\psi,\psi)
e^{-\int_0^\beta d\tau \(
\bm{\bar\psi}^T(\tau) \bm{Q} \bm{\psi}(\tau) +
\bm{\bar\psi}^T(\tau) \bm{\eta}(\tau) +
\bm{\bar\eta}^T(\tau) \bm{\psi}(\tau)
	\) }=\\
&=
e^{\int_0^\beta d\tau
	\bm{\bar\eta}^T(\tau) \bm{Q}^{-1} \bm{\eta}(\tau) 	}
\int D(\bar\psi,\psi)
e^{-\int_0^\beta d\tau
	\bm{\bar\psi}^T(\tau) \bm{Q} \bm{\psi}(\tau)
	}
\end{eqnarray} }
The Green function appears as a result of differentiation
\be
\Bigg[\frac{1}{\cZ[\bar\eta,\eta]}
\frac{\de^2\cZ[\bar\eta,\eta]}{\de \eta_x(\tau_1) \de \bar\eta_y(\tau_2)} \Bigg]_{\eta=\bar\eta=0}=\({\bf Q}^{-1}\)_{yx}\de\(\tau_1-\tau_2\) \equiv {\bf G}_{yx}\(\tau_1,\tau_2\)
\ee
It obeys equation
\be
{\bf Q} \^{\bf G}(\tau,\tau')=\de (\tau - \tau')
\ee
In the basis of one - particle states it obtains the form:
\be
\sum_k\bra{i} {\bf Q}  \ket{k}\bra{k}\^{\bf G}\(\tau_1,\tau_2\)\ket{j}=
\sum_k {\bf Q}_{ik} {\bf G}_{kj}\(\tau_1,\tau_2\)=
\de_{ij}\de (\tau_1 - \tau_2)
\ee
and
\be
G_{yx}\(\tau_1,\tau_2\)=\Bigg[\frac{1}{\cZ[\bar\eta,\eta]}
\frac{\de^2\cZ[\bar\eta,\eta]}
{\de \eta_x(\tau_1) 	\de \bar\eta_y(\tau_2)} \Bigg]_{\eta=\bar\eta=0}=
\frac{1}{\cZ}
\int D(\bar\psi,\psi) \psi_y(\tau_2) \bar\psi_x(\tau_1)
e^{-\int_0^\beta d\tau'
	\bm{\bar\psi}^T(\tau') \bm{Q} \bm{\psi}(\tau') }
\ee
recalling that $\braket{\la|\psi}=\psi_\la$ and $\braket{\psi|\la}=\bar\psi_\la$, the Green's "operator" may be defined as
\be
{\^G}\(\tau_1,\tau_2\)=
\frac{1}{\cZ}
\int D(\bar\psi,\psi)
\ket{\psi(\tau_2)} \bra{\psi(\tau_1)}  e^{-S[\bar\psi,\psi]}
\ee
and, the Green's function is given by matrix elements of this operator
\be
{\bf G}_{x,y}\(\tau_1,\tau_2\)=\bra{x} {\^G}\(\tau_1,\tau_2\) \ket{y}
\ee
Notice that these are matrix elements of $\hat{G}$ between the one - particle states, while  $\hat G$ is defined as an operator in Fock space. We will also define the restriction of $\hat{G}$ on the subspace of one - particle states:
$$
\hat{\bf G}=\sum_{x,y}\ket{x}\bra{x} {\^G}\(\tau_1,\tau_2\) \ket{y}\bra{y}
$$

Then
	\be
	\frac{\de \cZ}{\cZ}&=
	\cZ^{-1}\int D\(\bar\psi,\psi\)\(-\de S\) e^{-S[\bar\psi,\psi]}=
	-\cZ^{-1}\int D\(\bar\psi,\psi\)
	\int_0^\beta d\tau \Tr\(
	\de \^{\bf Q}
	\Ket{\psi(\tau)}
	\Bra{\psi(\tau)}
	\)
	e^{-S[\bar\psi,\psi]}=\\
	&=-\int_0^\beta d\tau \Tr
	\[\de \^{\bf Q}
	\(
	\cZ^{-1}\int D\(\bar\psi,\psi\)
	\Ket{\psi(\tau)}
	\Bra{\psi(\tau)}
	e^{-S[\bar\psi,\psi]}
	\)\]=\\&=
	-\int_0^\beta d\tau \Tr \[\de \^{\bf Q} \^{G}(\tau,\tau)\]=
	-\int_0^\beta d\tau \Tr \[\de \^{\bf Q} \^{\bf G}(\tau,\tau)\]
	\ee

\subsubsection{Matsubara frequencies representation}

We represent action as follows
\be
S[\bar\psi,\psi]=
\beta\sum_{\om_m}\sum_{ij} \bar\psi_i(\om_m) {\bf Q}_{ij}(\om_m) \psi_j(\om_m)=
\beta\sum_{\om_m} \bar{\bm\psi}^T(\om_m) \bm Q(\om_m) \bm{\psi}(\om_m)
\ee
The partition function receives the form
\be
\cZ[\bar\eta,\eta]
&=\int D(\bar\psi,\psi)
\exp\[
-\beta\sum_{\om_m}
\Big(
\bm{\bar\psi}^T(\om_m) \bm{Q}(\om_m) \bm{\psi}(\om_m) +
\bm{\bar\psi}^T(\om_m) \bm{\eta}(\om_m) +
\bm{\bar\eta}^T(\om_m) \bm{\psi}(\om_m)
\Big)
\]
=\\
&=
\e^{\beta\sum_{\om_m}
	\bm{\bar\eta}^T(\om_m) \bm{Q}^{-1}(\om_m) \bm{\eta}(\om_m) 	}
\int D(\bar\psi,\psi)
e^{-\beta\sum_{\om_m}
	\bm{\bar\psi}^T(\om_m) \bm{Q}(\om_m) \bm{\psi}(\om_m)
}
\ee
Therefore,
\be
{\bf G}_{ij}\(\om_m,\om_n\) &\equiv
\({\bf Q}^{-1}\)_{ij}\(\om_m\)\de_{mn}=
\Bigg[\frac{1}{\cZ[\bar\eta,\eta]}
\frac{\de^2\cZ[\bar\eta,\eta]}{\de \eta_j(\om_n) \de \bar\eta_i(\om_m)}
 \Bigg]_{\eta=\bar\eta=0}=\\
 &=\frac{\beta}{\cZ}\int D(\bar\psi,\psi)
\psi_i\(\om_m\) \bar\psi_j\(\om_n\)
 e^{-\beta\sum_{\om_m} 	\bm{\bar\psi}^T(\om_m) \bm{Q}(\om_m) \bm{\psi}(\om_m) }
\ee
and
\be
\^{ G}\(\om_m,\om_n\) =
\frac{\beta}{\cZ}\int D(\bar\psi,\psi)
\Ket{\psi\(\om_m\)} \Bra{\psi\(\om_n\)}
e^{-\beta\sum_{\om_m} 	\bm{\bar\psi}^T(\om_m) \bm{Q}(\om_m) \bm{\psi}(\om_m) }
\ee
with
\be
\sum_k {\bf Q}_{ik}\(\om_m\){\bf G}_{kj}\(\om_m,\om_n\)=\de_{ij}\de_{mn}
\ee
Also we have
\be
\^{\bf Q}\(\om_m\)\^G\(\om_m,\om_n\)=\de_{mn} \otimes 1_{1}
\ee
and
\be
\^{\bf Q}\(\om_m\)\^G\(\om_m,\om_m\)=1\otimes 1_{1}
\ee
Here $\otimes 1_{1}$ is the projector to the one - particle subspace of Fock space. In the following we will omit sometimes this projector if this does not cause ambiguities. Namely, we will write both
\be
\^{\bf Q}\(\om_m\)\^{\bf G}\(\om_m,\om_m\)=1\otimes 1_{1}
\ee
and
\be
\^{\bf Q}\(\om_m\)\^{\bf G}\(\om_m,\om_m\)={\bf 1}
\ee
Action may be written as
\be
S[\bar\psi,\psi]&=
\beta\sum_{\om_m}\Tr\(\^{\bf Q}(\om_m)\Ket{\psi(\om_m)}\Bra{\psi(\om_m)}\)=\beta\sum_{\om_m}\Tr\(\^{ Q}(\om_m)\kett{\psi(\om_m)}\bbra{\psi(\om_m)}\)
\ee
while
\be
\de S[\bar\psi,\psi]&=
\beta\sum_{\om_m}\Tr\(\de\^{\bf Q}(\om_m)\Ket{\psi(\om_m)}\Bra{\psi(\om_m)}\)
\ee
Variation of partition function reads
\be
\frac{\de \cZ}{\cZ}&=
\cZ^{-1}\int D\(\bar\psi,\psi\)\(-\de S\) e^{-S[\bar\psi,\psi]}=\\
&=-\beta\cZ^{-1}\int D\(\bar\psi,\psi\)
\sum_{\om_m}\Tr\(
\de \^{\bf Q}{(\om_m)}
\Ket{\psi(\om_m)}
\Bra{\psi(\om_m)}
\)
e^{-S[\bar\psi,\psi]}=\\
&=-\sum_{\om_m}\Tr \de \^{\bf Q}(\om_m)
\(
\beta\cZ^{-1}\int D\(\bar\psi,\psi\)
\Ket{\psi(\om_m)}
\Bra{\psi(\om_m)}
e^{-S[\bar\psi,\psi]}
\)=\\
&=-\sum_{\om_m}\Tr \de \^{\bf Q}(\om_m) \^{G}(\om_m,\om_m)
\ee

\section{ Tight Binding approximation - Fermions on 2D lattice in the presence of constant magnetic field}
\label{SectionIII}
\subsection{Dirac operator - Q}

In this section we start from the standard field - theoretic representation of the tight - binding model for the two - dimensional material in the presence of external magnetic field. We will introduce the notion of magnetic Brillouin zones in a manner slightly different from the standard one. It will be use latter in this form to express Hall conductivity as a topological invariant expressed through the Green functions.

The simplest tight binding Hamiltonian (for rectangular $2D$ lattice) in quantum mechanics of one particle has the form
\be
\^H=-t\sum_x\sum_{j=1,2} \ket{x}\bra{x+e_j}+h.c.\label{TBH}
\ee
Here $e_j$ is the lattice vector connecting two adjacent points.
In the presence of electromagnetic field we should modify this Hamiltonian adding a gauge field
\be
\^H=-t\sum_x\sum_{j=1,2} \ket{x}e^{-ieaA_j(x)/\hb}\bra{x+e_j}+h.c.
\label{Hg}
\ee
In many particle systems, i.e. in quantum field theory the Hamiltonian receives the form
\be
\^H(a^\dag,a)=-t\sum_x\sum_{j=1,2} \(a^\dag_x a_{x+e_j}+h.c.\)
\ee
and in the presence of electromagnetic field:
\be
\^H(a^\dag,a)=-t\sum_x\sum_{j=1,2} \(a^\dag_x e^{-ieaA_j(x)/\hb} a_{x+e_j}+h.c.\)  + \sum_x e a^\+_x A_0(x) a_x
\ee

{\it In the following we illustrate our derivation by direct consideration of this Hamiltonian. However, the obtained expressions for the electric current and conductivity are valid also for the Hamiltonian of a more general form, which may be checked easily at each step.} For the present consideration the lattice has to be rectangular. The generalization to the lattice of general form is not described here (although such a  generalization is straightforward).

We define translation operators
\be
\^T_j=\sum_x a^\+(x) e^{-ieaA_j(x)/\hb}a(x+e_j)
\ee
\be
\^T_j^\+=\sum_x a^\+(x+e_j) e^{ieaA_j(x)/\hb}a(x)=
\sum_x a^\+(x) e^{ieaA_j(x)/\hb}a(x-e_j)
\ee
Then the field Hamiltonian can be written as
\be
\^H=-t\sum_{j=1,2}\(\^T_j+\^T^\+_j\)
+ \sum_x e a^\+_x A_0(x) a_x
\ee
The Dirac operator in this construction is
\be
\^Q(a^\+,a)=\pd_\tau+
\^H(a^\dag,a)-\mu \^N(a^\dag,a)
\ee
with the fermion number operator
\be
N(a^\+,a)=\sum_i a^\+_i a_i
\ee
in Landau gauge $A_1=0$, $A_2=Bx_1$ we obtain the following expressions for translation operators
\be
\^T_1=\sum_x a^\+(x) a(x+e_1)
\tab
\^T_1^\+=\sum_x a^\+(x+e_1) a(x)= \sum_x a^\+(x) a(x-e_1)
\ee
\be
\^T_2=\sum_x a^\+(x) e^{-ieaA_2(x)/\hb}a(x+e_2)
=\sum_x a^\+(x) e^{-ieaBx_1/\hb}a(x+e_2)
\ee
\be
\^T_2^\+=\sum_x a^\+(x+e_2) e^{ieaBx_1/\hb} a(x)
\ee
Matrix elements of translation operators between the coherent states are
\be
\bbra{\psi}\^T_1\kett{\psi}=\sum_x \bar\psi(x)\psi(x+e_1) \tab
\bbra{\psi}\^T_1^\+\kett{\psi}=\sum_x \bar\psi(x+e_1)\psi(x)
\ee
\be
\bbra{\psi}\^T_2\kett{\psi}=\sum_x \bar\psi(x) e^{-ieaBx_1/\hb} \psi(x+e_2)
 \tab
\bbra{\psi}\^T_2^\+\kett{\psi}=\sum_x \bar\psi(x+e_2) e^{ieaBx_1/\hb} \psi(x)
\ee
In momentum space creation and annihilation operators are
\be
a^\+(x_1,x_2)&=
\int_{-\pi/a}^{\pi/a}dk_1 \int_{-\pi/a}^{\pi/a}dk_2
 a^\+(k_1,k_2) \braket{k_1,k_2|x_1,x_2}=\\
&=
\intl_{-\pi/a}^{\pi/a}dk_1 \intl_{-\pi/a}^{\pi/a}dk_2
a^\+(k_1,k_2) \frac{e^{-ik_1x_1-ik_2x_2}}{\sqrt{2\pi/a}}
\ee
\be
a(x_1,x_2)&=\int_{-\pi/a}^{\pi/a} dk_1 \int_{-\pi/a}^{\pi/a} dk_2
\braket{x_1,x_2|k_1,k_2}a(k_1,k_2)=\\
&=\intl_{-\pi/a}^{\pi/a} dk_1 \intl_{-\pi/a}^{\pi/a} dk_2
\frac{e^{ik_1x_1+ik_2x_2}}{\sqrt{2\pi/a}}a(k_1,k_2)
\ee
Through these operators the translation operators are expressed as
\be
\^T_1&=\sum_x
\intl_{-\pi/a}^{\pi/a}dk_1 \intl_{-\pi/a}^{\pi/a}dk_2
a^\+(k_1,k_2) \frac{e^{-ik_1x_1-ik_2x_2}}{\sqrt{2\pi/a}}
\intl_{-\pi/a}^{\pi/a} dk'_1 \intl_{-\pi/a}^{\pi/a} dk'_2
\frac{e^{ik'_1(x_1+a)+ik'_2x_2}}{\sqrt{2\pi/a}}a(k'_1,k'_2)\\
\ee
and
\be
\^T_1&=
\intl_{-\pi/a}^{\pi/a}dk_1 \intl_{-\pi/a}^{\pi/a}dk_2
a^\+(k_1,k_2) e^{ik_1a} a(k_1,k_2)
\ee

\be
\^T_1^\+&=
\intl_{-\pi/a}^{\pi/a}dk_1 \intl_{-\pi/a}^{\pi/a}dk_2
a^\+(k_1,k_2) e^{-ik_1a} a(k_1,k_2)
\ee
In the same way
\be
\^T_2&=\sum_x
\intl_{-\pi/a}^{\pi/a}dk_1 \intl_{-\pi/a}^{\pi/a}dk_2
a^\+(k_1,k_2) \frac{e^{-ik_1x_1-ik_2x_2}}{\sqrt{2\pi/a}}
e^{-i\frac{eaB}{\hb}x_1}
\intl_{-\pi/a}^{\pi/a} dk'_1 \intl_{-\pi/a}^{\pi/a} dk'_2
\frac{e^{ik'_1x_1+ik'_2(x_2+a)}}{\sqrt{2\pi/a}}a(k'_1,k'_2)\\
\ee
and
\be
\^T_2&=
\intl_{-\pi/a}^{\pi/a}dk_1 \intl_{-\pi/a}^{\pi/a}dk_2
a^\+(k_1,k_2) e^{ik_2a} a\(k_1+\frac{eaB}{\hb},k_2\)
\ee

\be
\^T_2^\+&=
\intl_{-\pi/a}^{\pi/a}dk_1 \intl_{-\pi/a}^{\pi/a}dk_2
a^\+(k_1,k_2) e^{-ik_2a} a\(k_1-\frac{eaB}{\hb},k_2\)
\ee
In the more simple case of a 1D system we have
\be
& \sum_x a^\+(x) x a(x) =
\sum_x \intl_{-\pi/a}^{\pi/a} dk  \intl_{-\pi/a}^{\pi/a} dk'
a^\+(k) \braket{k|x} x \braket{x|k'} a(k')=\\
&=\sum_x \intl_{-\pi/a}^{\pi/a} dk  \intl_{-\pi/a}^{\pi/a} dk'
a^\+(k) i\pd_k \braket{k|x}  \braket{x|k'} a(k')=\\
&= \intl_{-\pi/a}^{\pi/a} dk  \intl_{-\pi/a}^{\pi/a} dk'
a^\+(k) i\pd_k \braket{k|k'} a(k')=\\
&= \intl_{-\pi/a}^{\pi/a} dk
a^\+(k) i\pd_k  a(k)
\ee
Hence, in 2D in case of uniform electric field with
$
A_0(x)=-E_1 x_1-E_2 x_2
$
we obatain
\be
& \sum_x a^\+(x_1,x_2) \Big(-E_1 x_1-E_2 x_2\Big) a(x_1,x_2) =\\
&= \intl_{-\pi/a}^{\pi/a} dk_1 \intl_{-\pi/a}^{\pi/a} dk_2
a^\+(k_1,k_2) \Big(-E_1 i\pd_{k_1}-E_2 i\pd_{k_2}\Big)  a(k_1,k_2)
\ee

\subsection{Thermal average of the current}

Thermal average of electric current can be calculated as follows
\be
\braket{\^J_i(x)}
&= \frac{1}{\cZ}
\int D(\bar\psi,\psi)
\Bigg[\frac{1}{\beta}\int_0^\beta d\tau J_i(x)\(\bar\psi(\tau),\psi(\tau) \) \Bigg]
e^{-\int_0^\beta d\tau \Big(\bar\psi(\tau)\pd_\tau\psi(\tau)+H(\bar\psi(\tau),\psi(\tau))-\mu N(\bar\psi(\tau),\psi(\tau))\Big)}=\\
&= \frac{1}{\cZ}
\int D(\bar\psi,\psi)
\Bigg[\frac{1}{\beta}\int_0^\beta d\tau \(-\frac{\de H}{\de A_i(x)} \) \Bigg]
e^{-\int_0^\beta d\tau \Big(\bar\psi(\tau)\pd_\tau\psi(\tau)+H(\bar\psi(\tau),\psi(\tau))-\mu N(\bar\psi(\tau),\psi(\tau))\Big)}=\\
&= \frac{1}{\cZ}
\int D(\bar\psi,\psi)
\(-\frac{1}{\beta}\frac{\de S}{\de A_i(x)}\)
e^{-\int_0^\beta d\tau \Big(\bar\psi(\tau)\pd_\tau\psi(\tau)+H(\bar\psi(\tau),\psi(\tau))-\mu N(\bar\psi(\tau),\psi(\tau))\Big)}=\\
&= \frac{1}{\cZ} \frac{1}{\beta} \frac{\de \cZ}{\de A_i(x)}=
\frac{1}{\beta} \frac{\de \log \cZ}{\de A_i(x)}
\ee
Here we used expression for the current operator in the field systems described in Appendix \ref{AppendixD}.
Current averaged over the system area is
\be
\braket{\^{\bar J}_i} =
\frac{1}{L^2}\sum_x \braket{\^J_i(x)} =
\frac{1}{\cZ} \frac{1}{\beta L^2} \frac{\de \cZ}{\de A_i}=
\frac{1}{\beta L^2} \frac{\de \log \cZ}{\de A_i}
\ee
since
\be
\frac{\de \cZ}{\cZ}&=
-\int_0^\beta d\tau \Tr \[ \de \^{\bf Q} \^{G}(\tau)\] =
-\sum_{\om_m}\Tr \[\de \^{\bf Q}(\om_m) \^{G}(\om_m,\om_m)\]
%=-\int_0^\beta d\tau
%\Tr \[\(\pd_{A_i} \^{Q}\) \^{G}(\tau) \] \de A_i
\ee
we have
\be
\braket{\^ J_i(x)} =
-\frac{1}{\beta}\int_0^\beta d\tau
\Tr\[\frac{\de\^{\bf Q}}{\de A_i(x)} \^{G}(\tau) \]=
-\frac{1}{\beta}
\sum_{\om_m}\Tr \[\frac{\de \^{\bf Q}(\om_m)}{\de A_i(x)} \^{G}(\om_m,\om_m)\]
\ee
and
\be
\braket{\^{\bar J}_i} =
-\frac{1}{\beta L^2}\int_0^\beta d\tau
\Tr\[\frac{\de\^{\bf Q}}{\de A_i} \^{G}(\tau) \]=
-\frac{1}{\beta L^2}
\sum_{\om_m}\Tr \[\frac{\de \^{\bf Q}(\om_m)}{\de A_i} \^{G}(\om_m,\om_m)\]
\ee
where $\de A_i$ is homogeneous.

\subsection{Magnetic Brillouin zones (MBZ)}
\subsubsection{Single particle completeness relation }
For the single particle 1D lattice we have in momentum space
\be
1=\intl_{\De}^{\De+\frac{2\pi}{a}} dk \ket{k}\bra{k}=
\sum_{n=0}^{N-1}
\int_{\De+n\frac{2\pi}{Na}}^{\De+(n+1)\frac{2\pi}{Na}} dk_n \ket{k_n}\bra{k_n}
\ee
where Brillouin zone may be defined as
\be k\in \[\De,\De+\frac{2\pi}{a}\] = BZ\ee
with arbitrary $\Delta$. Correspondingly,
\be k_n\in \[\De+n\frac{2\pi}{Na},\De+ (n+1) \frac{2\pi}{Na}\]\ee

We may define
$k_n =\ka +n\frac{2\pi}{aN}$ where
$\ka\in \[\De,\De+\frac{2\pi}{Na}\]\equiv\frac{BZ}{N}$
\be
1=
\sum_{n=0}^{N-1}\int_{\De}^{\De+\frac{2\pi}{Na}} d\ka
\Ket{\ka +n\frac{2\pi}{Na}}
\Bra{\ka +n\frac{2\pi}{Na}}
\ee
The inner product
\be
\Braket{\ka +n\frac{2\pi}{Na}|\ka' +n'\frac{2\pi}{Na}}
\ee
is nonzero for
\be
\(\ka-\ka'\)=(n-n')\frac{2\pi}{Na}
\ee
But, since $\ka\in \[\De,\De+\frac{2\pi}{Na}\]$, this means that $\ka=\ka'$ and $n=n'$ separately. Therefore, we denote the states $\ket{k_n}=\ket{\ka,n}$, for which
\be
\braket{\ka,n|\ka',n'}=\de(\ka-\ka')\de_{nn'}
\ee
hence
\be
1=
\sum_{n=0}^{N-1}\int_{\De }^{\De +\frac{2\pi}{Na}} d\ka \ket{\ka,n}\bra{\ka,n}=
\sum_{n=0}^{N-1}\int_{\frac{BZ}{N}}  d\ka \ket{\ka,n}\bra{\ka,n}
\ee
Further, we have
\be
&\bra{p,n}a^\+(k) a(k')\ket{p',n'}=
\bra{p,n}a^\+(\ka,m) a(\ka',m')\ket{p',n'}=
\delta(p-\ka) \de_{mn} \delta(p'-\ka') \de_{m'n'}
\ee
where $k,k'\in BZ$, and $p,p',\ka,\ka'\in BZ/N$,  \,\,\, $n,m,n',m'=1,...,N$\\

Discrete momentum corresponding to any value of $n$ may be represented as
\be
\frac{2\pi}{Na}n=\frac{2\pi}{Na}\nu m -M\frac{2\pi}{a}
\label{nm}
\ee
where $n,m=1,...,N$ and $\nu,M\in  \N$, and $(\nu,N)=1$, i.e. $\nu$ and $N$ are mutually simple numbers.

Let us compose in this way two different values of $n$ through different values of $m$ and $M$ but with the same $\nu$ and $N$:
\be
n_1=m_1\nu-M_1N \tab
n_2=m_2\nu-M_2N
\ee
Now if we require that $n_1 = n_2$ then
\be
\frac{\nu}{N}(m_1-m_2)=(M_1-M_2)
\label{nnmm}
\ee
since $|m_1-m_2|<N$ and $(\nu,N)=1$, we see that (\ref{nnmm}) holds only for $m_1=m_2$, and  $M_1=M_2$. That means that for any specific $m$ there is only one specific $n$, and both are between 1 and $N$. Hence, the (\ref{nm}) for fixed $\nu$ and $N$ is a one to one transformation between $n$ and the pair $m,M$. Hence
\be
1=
\sum_{n=0}^{N-1}\int_{\De }^{\De+\frac{2\pi}{Na}} d\ka
\Ket{\ka +n\frac{2\pi}{aN}\nu }
\Bra{\ka +n\frac{2\pi}{aN}\nu }
\ee
the inner product
\be
\Braket{\ka +n\frac{2\pi}{aN}\nu|\ka' +n'\frac{2\pi}{aN}\nu}
\ee
is nonzero for
\be
\(\ka-\ka'\)=(n'-n)\frac{2\pi}{Na}\nu
\ee
but, since $\ka,\ka'\in  \[\De,\De+\frac{2\pi}{Na}\]$ and $\nu>1$, this means that $\ka=\ka'$ and $n=n'$, separately. Therefore, we may represent the states
\be \ket{k_n}=\ket{\ka +n\frac{2\pi}{aN}\nu}\equiv\ket{\ka,n}_\De
\label{k_n}
\ee
 for which
\be
\braket{\ka,n|\ka',n'}=\de(\ka-\ka')\de_{nn'}
\ee
hence
\be
1=
\sum_{n=0}^{N-1}\int_{\De }^{\De +\frac{2\pi}{Na}} d\ka
 \Ket{\ka +n\frac{2\pi}{aN}\nu} \Bra{\ka +n\frac{2\pi}{aN}\nu}=
\sum_{n=0}^{N-1}\int_{\frac{BZ}{N}}  d\ka \ket{\ka,n}\bra{\ka,n}
\ee

The similar decomposition of completeness relation in 2D reads
\be
1&=
\sum_{n=0}^{N-1}
\int_{\De}^{\De +\frac{2\pi}{Na}} du
\int_{-\frac{\pi}{a} }^{\frac{\pi}{a}} dv
\Ket{u +n\frac{2\pi}{aN}\nu,v} \Bra{u +n\frac{2\pi}{aN}\nu,v}=\\
&=\sum_{n=0}^{N-1}
\int_{\De}^{\De +\frac{2\pi}{Na}} du
\int_{-\frac{\pi}{a} }^{\frac{\pi}{a}} dv
\ket{un,v}\bra{un,v} =\\
&=\sum_{n=0}^{N-1}
\int_{0}^{\frac{2\pi}{a}} du
\int_{0}^{\frac{2\pi}{a}} dv
\Ket{u +n\frac{2\pi}{aN}\nu,v} \Bra{u +n\frac{2\pi}{aN}\nu,v}
\te\(u-\De\)\te\(\De+\frac{2\pi}{Na}-u\)
\ee
 where
 \be
 \ket{un,v}\equiv  \Ket{u +n\frac{2\pi}{aN}\nu,v}
 \ee

\subsubsection{Dirac operator matrix elements - $\bra{\ka,n}\^Q\ket{\ka',n'}$}

Dirac operator  Q has the form
\be
\^Q(a^\+,a)=\pd_\tau+
\^H(a^\dag,a)-\mu \^N(a^\dag,a)
\ee
Let us consider first for simplicity the 1D model, where we have
\be
N(a^\+,a)=\sum_x a^\+_x a_x=\int_{BZ} dk a^\+_k a_k=
\sum_{n=1}^N \int_{\frac{BZ}{N}} d\ka a^\+_{\ka,n} a_{\ka,n}
\ee
and
\be
\bra{p,n}\^N\ket{p',n'}&=
\int_{BZ} dk \bra{p,n} a^\+_k a_k \ket{p',n'}=
\sum_{m=1}^N \int_{\frac{BZ}{N}} dk \bra{p,n} a^\+_{\ka,m} a_{\ka,m} \ket{p',n'}=\\
&=\sum_{m=1}^N \int_{\frac{BZ}{N}} dk \de(p-\ka)\de_{n,m}\de(p'-\ka)\de_{n',m}=\de(p-p')\de_{n,n'}
\ee
while
\be
&\bra{p,n} \sum_x a^\+(x) x a(x)  \ket{p',n'}=
\intl_{BZ} dk \bra{p,n}  a^\+(k) i\pd_k  a(k) \ket{p',n'}=\\
&= \intl_{BZ} dk \Big[ i\pd_{k'} \bra{p,n}  a^\+(k)   a(k') \ket{p',n'} \Big]_{k'=k}=
i\pd_{p'}\intl_{BZ} dk  \de((p,n)-k)  \de((p',n')-k) =\\
&= i\pd_{p'} \de((p,n)-(p',n'))= i\pd_{p'} \de(p-p') \de_{n,n'}
\ee

Now let us come back to 2D, where the notations are slightly more complicated. Namely, the basis vectors are $\ket{(u,n),v}$, where $u\in BZ/N$, $n=1,...,N$ and, $v\in BZ$.

For the matrix elements of the particle number operator
${\^N(a^\+,a)}$ we have
\be
\bra{(u,n),v}\^N\ket{(u',n'),v'}&=
\intl_{BZ}dk_1 \intl_{BZ}dk_2
\bra{(u,n),v} a^\+(k_1,k_2)  a(k_1,k_2) \ket{(u',n'),v'}=\\
&=\de(v-v')\de(u-u') \de_{nn'}
\ee

Matrix elements of operator
${\sum_x e a^\+_x A_0(x) a_x}$ read (in case of uniform electric field)
$
A_0(x)=-E_1 x_1-E_2 x_2
$
\be
&e\sum_x \bra{(u,n),v} a^\+(x_1,x_2) \Big(-E_1 x_1-E_2 x_2\Big) a(x_1,x_2) \ket{(u',n'),v'} =\\
&= e\(-E_1 i\pd_{u} -E_2 i\pd_{v}\) \de(u-u') \de_{n,n'} \de(v-v')
\ee

Matrix elements of the combination of translation operators
${\sum_{j=1,2}\(\^T_j+\^T^\+_j\) }$ can be represented as follows. Let us start from translation in the first direction:
\be
\bra{(u,n),v}\^T_1\ket{(u',n'),v'}&=
\intl_{BZ}dk_1 \intl_{BZ}dk_2
\bra{(u,n),v} a^\+(k_1,k_2) e^{ik_1a} a(k_1,k_2) \ket{(u',n'),v'}=\\
&=\de(v-v')e^{i(u,n)a}\de(u-u') \de_{nn'}
\ee
We obtain
\be
\bra{(u,n),v}\^T_1\ket{(u',n')v'}
&=\de(v-v')e^{i(u,n)a}\de(u-u') \de_{nn'}
\ee
Translation in the second direction reads
\be
\bra{(u,n),v}\^T_1^\+\ket{(u',n')v'}
&=\de(v-v')e^{-i(u,n)a}\de(u-u') \de_{nn'}
\ee
while
\be
\^T_2&=
\intl_{-\pi/a}^{\pi/a}dk_1 \intl_{-\pi/a}^{\pi/a}dk_2
a^\+(k_1,k_2) e^{ik_2a} a\(k_1+\frac{eaB}{\hb},k_2\)
\ee

\be
\^T_2^\+&=
\intl_{-\pi/a}^{\pi/a}dk_1 \intl_{-\pi/a}^{\pi/a}dk_2
a^\+(k_1,k_2) e^{-ik_2a} a\(k_1-\frac{eaB}{\hb},k_2\)
\ee

In accordance with Eq. (\ref{k_n}) let us require that magnetic field is quantized and has the form
\be
\frac{2\pi}{Na}\nu=\frac{eaB}{\hb}
\ee
with mutually simple integer numbers $\nu$ and $N$.
We define magnetic flux quantum $\Phi_0=\frac{\hb}{e}2\pi$ and $\Phi=a^2B$. This gives the following condition on the magnetic flux through the lattice cell
\be
\frac{\nu}{N}=\frac{\Phi}{\Phi_0}
\ee
In practise for any value of $B$ (given by a rational number times $\frac{h}{e a^2}$) we can choose $\nu$ and $N$ to fulfill the above relation.

Now we rewrite $\^T_2$ and $\^T_2^\+$ as follows
\be
\^T_2&=
\sum_{n=1}^N \intl_{-\pi/Na}^{\pi/Na}d\ka \intl_{-\pi/a}^{\pi/a}dk_2
a^\+((\ka,n),k_2) e^{ik_2a} a\((\ka,n+1),k_2\)
\ee

\be
\^T_2^\+&=
\sum_{n=1}^N \intl_{-\pi/Na}^{\pi/Na}d\ka \intl_{-\pi/a}^{\pi/a}dk_2
a^\+((\ka,n),k_2) e^{-ik_2a} a\((\ka,n-1),k_2\)
\ee
For the matrix elements we obtain
\be
\bra{(u,n),v}\^T_2 \ket{(u',n'),v'}&=
\sum_{m=1}^N \intl_{-\pi/Na}^{\pi/Na}d\ka \intl_{-\pi/a}^{\pi/a}dk_2
\bra{(u,n),v} a^\+((\ka,n),k_2) e^{i(\ka,m)a} a\((\ka,m+1),k_2\) \ket{(u',n'),v'}=\\
&=\de(v-v')e^{iva}\de(u-u') \de_{n,n'-1}
\ee
that is
\be
\bra{(u,n),v}\^T_2 \ket{(u',n'),v'}=\de(v-v')e^{iva}\de(u-u') \de_{n,n'-1}
\ee
and
\be
\bra{(u,n),v}\^T_2^\+ \ket{(u',n'),v'}=\de(v-v')e^{-iva}\de(u-u') \de_{n,n'+1}
\ee
As a result we have
\be
&\bra{(u,n),v}
\sum_{j=1,2}\(\^T_j+\^T^\+_j\)
\ket{(u',n'),v'}=\\
&\[\(e^{i(u,n)a} + e^{-i(u,n)a}\) \de_{nn'} + e^{iva} \de_{n,n'-1} +e^{-iva} \de_{n,n'+1} \] \de(u-u') \de(v-v')
\ee
In Matsubara frequency representation we obtain for the Dirac operator
\be
&Q_{unvu'n'v'}^{(m)}=\bra{(u,n),v}\^Q^{(m)} \ket{(u',n'),v'}=\\
&\de(v-v')\de(u-u') \de_{n,n'}
\Big[-i \om_m-\mu+e\(-E_1i\pd_u-E_2i\pd_v\)-2t\cos((u,n)a)
\Big]-\\
&-\de(v-v')\de(u-u') t \Big[ e^{iva} \de_{n,n'-1} + e^{-iva} \de_{n,n'+1}\Big]=\\
&=\de(v-v')\de(u-u')\Bigg[Q_{[0]}^{(\om_m)}\de_{nn'}+Q_{[-1]}\de_{n,n'-1}+Q_{[+1]}\de_{n,n'+1}\Bigg]=\\
&=\de(v-v')\de(u-u')Q^{(\om_m)}_{uvnn'}\label{Quv}
\ee
Here
\be
Q^{(\om_m)}_{uvnn'} = Q_{[0]}^{(\om_m)}\de_{nn'}+Q_{[-1]}\de_{n,n'-1}+Q_{[+1]}\de_{n,n'+1}\label{QHarper}
\ee
and we denote
\be
Q_{[0]}^{\om_m}(u,n)=
-i \om_m-\mu-2t\cos((u,n)a)
\ee
\be
Q_{[-1]}(v)=-t e^{iva} \tab
Q_{[+1]}(v)=-t e^{-iva}
\ee
and
\be
Q^{(\om_m)}_{uvnn'}=
\Big[-i \om_m-\mu-2t\cos((u,n)a)\Big]\de_{n,n'}-t
e^{iva}\de_{n,n'-1} -t
e^{-iva} \de_{n,n'+1}
\ee
with
\be
(u,n)a=\(u+n\frac{2\pi}{Na}\nu\)a=ua+n\frac{2\pi}{N}\nu
\ee
{\it Notice that these expressions for the function $Q^{(\om_m)}_{uvnn'}$ are specific for the tight - binding Hamiltonian of the form of Eq. (\ref{TBH}). However, the form of Eq. (\ref{Quv}) remains the same for the tight - binding Hamiltonian of general form defined on the rectangular lattice provided that the magnetic flux through the lattice cell is quantized in a considered above way.}

\subsection{Linear and quadratic response}
Response to external potential of electromagnetic field gives electric current. We will be interested in the electric current averaged over the system area. It appeas as the response to contant external electromagnetic potential. The response of the latter with respect to constant external electric field gives conductivity. Therefore, we consider the response of Dirac operator to both:
\be
&\^ {\bf Q}\ra \^ {\bf Q}'=\^ {\bf Q}+\de^{\vec E} \^ {\bf Q}=\^ {\bf Q}+\frac{\partial \^ {\bf Q}}{\partial E_i}E_i \\
&\^ {\bf G}\ra \^ {\bf G}'=\^ {\bf G}+\de^{\vec E} \^ {\bf G}=\^ {\bf G}+\frac{\partial \^ {\bf G}}{\partial E_i}E_i
\ee
The quadratic response reads
\be
\frac{\de\^{\bf {\bf Q}}'}{\partial A_i} \^{{\bf G}}'&=
\(\frac{\partial\^{{\bf Q}}}{\partial A_i}+ \frac{\partial^2 \^{{\bf Q}}}{\partial A_i\de E_j}E_j\)
\(\^{{\bf G}}+\de^{\vec E}\^ {\bf G}\)=\\
&=\frac{\partial \^{\bf {\bf Q}}}{\partial A_i} \^{{\bf G}}+ \frac{\partial \^{\bf {\bf Q}}}{\partial A_i}\de^{\vec E}\^ {\bf G}
\ee
Since
\be
\frac{\partial^2 \^{{\bf Q}}}{\partial A_i\partial E_j}=0
\ee
we obtain
\be
\^ {\bf {\bf Q}}' \^ {\bf G}'=
\(\^ {\bf {\bf Q}}+\de^{\vec E}\^ {\bf {\bf Q}}\)\(\^ {\bf G}+\de^{\vec E} \^ {\bf G}\)=
\^ {\bf {\bf Q}} \^ {\bf G}={\bf 1}
\ee
up to the first order, and
\be
\de^{\vec E}\^ {\bf G}=-\^ {\bf G} \de^{\vec E}\^ {\bf Q} \^ {\bf G}=
-\^ {\bf G} \frac{\partial \^ {\bf Q}}{\partial E_i} \^ {\bf G} E_i
\ee
Up to the second order we have
\be
\frac{\partial \^{{\bf Q}}'}{\partial  A_i} \^{{\bf G}}'=
\frac{\partial \^{{\bf Q}}}{\partial  A_i} \^{{\bf G}}-
\frac{\partial \^{{\bf Q}}}{\partial  A_i}\^ {\bf G} \frac{\partial  \^ {\bf Q}}{\partial  E_j} \^ {\bf G} E_j
\ee
This gives
\be
\braket{\^ J_i(x)}_{\vec E} =
\frac{1}{\beta}\sum_{\om_m}
\Tr\[\frac{\partial \^{{\bf Q}}(\om_m)}{\partial  A_i}\^ {\bf G}(\om_m) \frac{\partial  \^ {\bf Q}(\om_m)}{\partial  E_j} \^ {\bf G}(\om_m) \]E_j
\equiv \si_{ij} E_j
\ee
the conductivity summed over the system area is
\be
\si_{ij}= \frac{1}{\beta}\int_0^\beta d\tau
\Tr\[\frac{\partial \^{{\bf Q}}}{\partial  A_i}\^ {\bf G}\(\tau\) \frac{\partial  \^ {\bf Q}}{\partial  E_j} \^ {\bf G}\(\tau\) \]=
\frac{1}{\beta} \sum_{\om_m}
\Tr\[\frac{\partial \^{{\bf Q}}{(\om)}}{\partial  A_i}\^ {\bf G}{(\om)} \frac{\partial  \^ {\bf Q}\(\om\)}{\partial  E_j} \^ {\bf G}\(\om\) \]
\ee
Current averaged over the system area is
\be
\braket{\^ {\bar J}_i}_{\vec E} =
\frac{1}{\beta L^2} \sum_{\om_m}
\Tr\[\frac{\partial \^{{\bf Q}}(\om_m)}{\partial  A_i}\^ {\bf G}(\om_m) \frac{\partial  \^ {\bf Q}(\om_m)}{\partial  E_j} \^ {\bf G}(\om_m) \]E_j
\equiv \bar \si_{ij} E_j
\ee
And the average conductivity is given by
\be
\bar \si_{ij}= \frac{1}{\beta L^2}\int_0^\beta d\tau
\Tr\[\frac{\partial \^{{\bf Q}}}{\partial  A_i}\^ {\bf G}\(\tau\) \frac{\partial  \^ {\bf Q}}{\partial  E_j} \^ {\bf G}\(\tau\) \]=
\frac{1}{\beta L^2} \sum_{\om_m}
\Tr\[\frac{\partial \^{{\bf Q}}{(\om)}}{\partial  A_i}\^ {\bf G}{(\om)} \frac{\partial  \^ {\bf Q}\(\om\)}{\partial  E_j} \^ {\bf G}\(\om\) \]
\ee

Using the MBZ (magnetic Brillouin zone) decomposition, we obtain:
\be
&\Tr\[\frac{\partial \^{{\bf Q}}{(\om)}}{\partial  A_i}\^ {\bf G}{(\om)} \frac{\partial  \^ {\bf Q}\(\om\)}{\partial  E_j} \^ {\bf G}\(\om\) \]=
\sum_n \intl_{\Delta }^{\Delta +\frac{2\pi}{Na}} dp_1 \intl_{BZ} dp_2
\bra{np_1p_2}
\frac{\partial \^{{\bf Q}}{(\om)}}{\partial  A_i}\^ {\bf G}{(\om)} \frac{\partial  \^ {\bf Q}\(\om\)}{\partial  E_j} \^ {\bf G}\(\om\)
\ket{np_1p_2}=\\
&=\sum_{nn'n''n'''}
\intl_{\Delta }^{\Delta +\frac{2\pi}{Na}} dp_1
\intl_{\Delta }^{\Delta +\frac{2\pi}{Na}} dp'_1
\intl_{\Delta }^{\Delta +\frac{2\pi}{Na}} dp''_1
\intl_{\Delta }^{\Delta +\frac{2\pi}{Na}} dp'''_1
\intl_{BZ} dp_2
\intl_{BZ} dp'_2
\intl_{BZ} dp''_2
\intl_{BZ} dp'''_2\\
&\Bra{np_1p_2}
\frac{\partial \^{{\bf Q}}{(\om)}}{\partial  A_i}
\Ket{n'p'_1p'_2}\bra{n'p'_1p'_2}
\^ {\bf G}{(\om)}
\Ket{n''p''_1p''_2}\bra{n''p''_1p''_2}
\frac{\partial  \^ {\bf Q}\(\om\)}{\partial  E_j}
\Ket{n'''p'''_1p'''_2}\bra{n'''p'''_1p'''_2}
\^ {\bf G}\(\om\)
\ket{np_1p_2}
\ee
Recall that
\be
&\bra{(u,n),v} \sum_x e a^\+_x A_0(x) a_x\ket{(u',n'),v'}=\\
&=e\sum_x \bra{(u,n),v} a^\+(x_1,x_2) \Big(-E_1 x_1-E_2 x_2\Big) a(x_1,x_2) \ket{(u',n'),v'} =\\
&=e\bra{(u,n),v} \int dp_1 \int dp_2 a^\+(p_1,p_2)
\(-E_1i\pd_{p_1}-E_2i\pd_{p_2}\) \ket{(u',n'),v'}=\\
&= e\(-E_1 i\pd_{u} -E_2 i\pd_{v}\) \partial (u-u') \partial _{n,n'} \partial (v-v')
\ee
Then
\be
&\bra{np_1p_2}
\frac{\partial  \^ {\bf Q}\(\om\)}{\partial  E_j}
\Ket{n'p'_1p'_2}=
\bra{np_1p_2}
\frac{\partial  }{\partial  E_j} \sum_x e a^\+_x A_0(x) a_x
\Ket{n'p'_1p'_2}=
-ei\pd_{p_j}\partial _{nn'}\partial (p_1-p'_1)\partial (p_2-p'_2)
\ee
and
\be
&\bra{np_1p_2}
\frac{\partial  \^ {\bf Q}\(\om\)}{\partial  A_i}
\Ket{n'p'_1p'_2}=
   \frac{e}{\hbar} \delta(p_1-p_1')\delta(p_2-p_2') \partial_{p_i}Q^{\om_m}_{p_1p_2nn'}
\ee
By
${\bf Q}^{\om_m}_{p_1p_2} $ we denote $N\times N$ matrix with elements $Q^{\om_m}_{p_1p_2nn'} $, by Tr$_n$ we represent the trace with respect to indices $n$.

As a result we represent
\be
&\Tr\[\frac{\partial \^{{\bf Q}}{(\om)}}{\partial  A_i}\^ {\bf G}{(\om)} \frac{\partial  \^ {\bf Q}\(\om\)}{\partial  E_j} \^ {\bf G}\(\om\) \]
=\sum_{nn'n''n'''} \intl_{\Delta }^{\Delta  +\frac{2\pi}{Na}} dp_1 \intl_{BZ} dp_2\\
&\Bra{np_1p_2}
\frac{\partial \^{{\bf Q}}{(\om)}}{\partial  A_i}
\Ket{n'p_1p_2}\bra{n'p_1p_2}
\^ G{(\om)}
\Ket{n''p_1p_2}\bra{n''p_1p_2}\(-ei\pd_{p_j}\)
\Ket{n'''p_1p_2}\bra{n'''p_1p_2}
\^ G\(\om\)
\ket{np_1p_2}=\\
&=- \frac{e^2}{\hbar} i\intl_{\Delta }^{\Delta  +\frac{2\pi}{Na}} dp_1 \intl_{BZ} dp_2
\Tr_n\Bigg[
\frac{\pd\bm{Q}_{p_1p_2}^{(\om)}}{\pd p_i}
\bm{G}_{p_1p_2}^{(\om)}
\frac{\pd\bm{G}_{p_1p_2}^{(\om)}}{\pd p_j}
\Bigg] \delta(p_1-p_1)\delta(p_2-p_2)=\\
&= \frac{e^2}{\hbar} i\intl_{\Delta }^{\Delta  +\frac{2\pi}{Na}} dp_1 \intl_{BZ} dp_2
\Tr_n\Bigg[
\frac{\pd\bm{Q}_{p_1p_2,}^{(\om)}}{\pd p_i}
\bm{G}_{p_1p_2}^{(\om)}
\bm{G}_{p_1p_2}^{(\om)}
\frac{\pd\bm{Q}_{p_1p_2}^{(\om)}}{\pd p_j}
\bm{G}_{p_1p_2}^{(\om)}
\Bigg]\delta(p_1-p_1)\delta(p_2-p_2) =\\
&= - \frac{e^2}{\hbar} \intl_{\Delta }^{\Delta  +\frac{2\pi }{Na}} dp_1 \intl_{BZ} dp_2
\Tr_n\Bigg[
\frac{\pd\bm{Q}_{p_1p_2}^{(\om)}}{\pd p_i}
\bm{G}_{p_1p_2}^{(\om)}
\frac{\pd\bm{Q}_{p_1p_2}^{(\om)}}{\pd \om}
\bm{G}_{p_1p_2}^{(\om)}
\frac{\pd\bm{Q}_{p_1p_2}^{(\om)}}{\pd p_j}
\bm{G}_{p_1p_2}^{(\om)}
\Bigg]\delta(p_1-p_1)\delta(p_2-p_2)\label{finD}
\ee
Here delta function of zero momentum should be understood as
$$\delta(p_1-p_1)\delta(p_2-p_2) = \frac{a^2}{(2\pi)^2} \sum_{x} e^{i(p - p )x} = \frac{L^2}{(2\pi)^2}  $$
where $L^2$ is the system area.
We also used above that
\be
\frac{\partial \bm{Q}_{p_1p_2}^{(\om)}}{\partial  \om} =-i
\ee
One can easily check using the property of trace that Eq. (\ref{finD}) is not changed if we replace $\Delta$ by $\Delta + \frac{2 \pi}{Na}m$ with any integer $m$.
Therefore, we can extend the integration to the whole Brillouin zone and divide the resulting result by $N$:
\be
&\Tr\[\frac{\partial \^{\bf Q}{(\om)}}{\partial  A_i}\^ {\bf G}{(\om)} \frac{\partial  \^ {\bf Q}\(\om\)}{\partial  E_j} \^ {\bf G}\(\om\) \]=\\
&-\frac{L^2}{N}
\frac{e^2}{\hb}\intl_{BZ} \frac{dp_1}{2\pi}  \frac{dp_2}{2\pi}
\Tr_n\Bigg[
\frac{\partial \bm{Q}_{p_1p_2}^{(\om)}}{\partial  p_i}
\bm{G}_{p_1p_2}^{(\om)}
\frac{\partial \bm{Q}_{p_1p_2}^{(\om)}}{\partial  \om}
\bm{G}_{p_1p_2}^{(\om)}
\frac{\partial \bm{Q}_{p_1p_2}^{(\om)}}{\partial  p_j}
\bm{G}_{p_1p_2}^{(\om)}
\Bigg]
\ee

\subsection{Final expression for average conductivity}

The conductivity averaged over the system area is given by
\be
\bar \si_{ij}&=
\frac{1}{\beta L^2} \sum_{\om_m}
\Tr\[\frac{\de\^{\bm Q}{(\om)}}{\de A_i}\^ {\bm G}(\om) \frac{\de \^ {\bm Q}\(\om\)}{\de E_j} \^ {\bm G}\(\om\) \]
\ee

At low temperature $T\ra 0$ the sum may be replaced by an integral $\sum_{\om_m}\ra \frac{1}{2\pi T}\int d\om$
\be
\bar \si_{ij}&=
-\frac{e^2}{2\pi \hb  4\pi^2 N}
\int d\om \intl_{BZ} dp_1 \intl_{BZ} dp_2 \Tr
\[
\frac{\pd {\bm Q}_{p_1p_2}^\om}{\pd p_i}
{\bm G}_{p_1p_2}^\om
\frac{\pd {\bm Q}_{p_1p_2}^\om}{\pd \om}
{\bm G}_{p_1p_2}^\om
\frac{\pd {\bm Q}_{p_1p_2}^\om}{\pd p_j}
{\bm G}_{p_1p_2}^\om
\]=\\
&=-\frac{e^2}{2\pi \hb 4 \pi^2 N }
\int d\om \intl_{BZ} dp_1 \intl_{BZ} dp_2
\sum_{a,...,f=1}^N
\[
\frac{\pd Q_{p_1p_2ab}^\om}{\pd p_i}
G_{p_1p_2bc}^\om
\frac{\pd Q_{p_1p_2cd}^\om}{\pd \om}
G_{p_1p_2de}^\om
\frac{\pd Q_{p_1p_2ef}^\om}{\pd p_j}
G_{p_1p_2fa}^\om
\]\label{sfin}
\ee

where
\be
Q^{\om_m}_{uvnn'}=
\Big[-i \om_m-\mu-2t\cos((u,n)a)\Big]\de_{n,n'}-t
e^{iva}\de_{n,n'-1} -t
e^{-iva} \de_{n,n'+1}
\ee
while
\be
(u,n)a=\(u+n\frac{2\pi}{Na}\nu\)a=ua+n\frac{2\pi}{N}\nu
\ee
i.e.
\be
Q^{\om_m}_{p_1p_2nn'}=
\[-i\om_m-\mu-2t\cos\(p_1a+n\frac{2\pi}{N}\nu\)\]\de_{n,n'}-t
e^{ip_2a}\de_{n,n'-1} -t
e^{-ip_2a} \de_{n,n'+1}
\ee
with
\be
\frac{\pd Q^{\om_m}_{p_1p_2nn'} }{\pd \om }=-i\de_{nn'}
\ee
and
\be
\frac{\pd Q^{\om_m}_{p_1p_2nn'} }{\pd p_1 }=\de_{nn'}
2at\sin\(p_1a+n\frac{2\pi}{N}\nu\)
\ee
\be
\frac{\pd Q^{\om_m}_{p_1p_2nn'} }{\pd p_2 }=
-itae^{ip_2a}\de_{n,n'-1}+itae^{-ip_2a} \de_{n,n'+1}
\ee

For the antisymmetric (Hall) part of conductivity we obtain:
\be
\boxed{
\bar \si^{AS}_{ij}=
\epsilon_{ij}\, \frac{e^2}{h} \, \frac{1}{N}\, \frac{\epsilon^{abc}}{24  \pi^2} \,
\int d\om \intl_{BZ} dp_1  dp_2 \Tr
\Bigl[
\frac{\pd {\bm Q}_{p_1p_2}^\om}{\pd p_a}
{\bm G}_{p_1p_2}^\om
\frac{\pd {\bm Q}_{p_1p_2}^\om}{\pd p_b}
{\bm G}_{p_1p_2}^\om
\frac{\pd {\bm Q}_{p_1p_2}^\om}{\pd p_c}
{\bm G}_{p_1p_2}^\om
\Bigr]} \label{HC}
\ee

{\bf This is the main result of our paper. It is valid for the non - interacting tight - binding fermionic system of general form although the derivation has been illustrated by consideration of the model with the Hamiltonian of Eq. (\ref{TBH}).} This expression has been derived for the system defined on rectangular lattice with the magnetic flux through the elementary lattice cell given by a rational number $\nu/N$ times elementary magnetic flux. One can see that the Hall conductivity may be represented in the form of the product
$$
\sigma_H = \frac{e^2}{h} \, \frac{1}{N}\,{\cal N}
$$
where $\cal N$ is the integer topological invariant composed of the $N\times N$ matrices $\bf Q$. The proof that this expression is a topological invariant is given in Appendix \ref{AppendixE}. For the non - interacting systems this invariant {\it should} be equal to the integer multiple of $N$ in order to provide the integer QHE.

We expect that Eq. (\ref{HC}) remains valid also for the case of an interacting system, where matrix $\bf G$ is replaced by the complete propagator defined in the magnetic Brillouin zone. In this case the value of $\cal N$ is already not necessarily equal to the integer multiple of $N$. This way we may obtain the topological description of the fractional quantum Hall effect.

\section{Numerical results}
\label{SectionIV}
\subsection{The considered system. Diophantine equation.}

In this section we illustrate the general expressions obtained above by numerical results obtained for the system with the one - particle Hamiltonian of the form
\be
\^H(a^\dag,a)=v_1\sum_x \(a^\dag_x e^{-ieaA_1(x)/\hb} a_{x+e_1}+h.c.\)  + v_2\sum_x \(a^\dag_x e^{-ieaA_2(x)/\hb} a_{x+e_2}+h.c.\)+  \sum_x e a^\+_x A_0(x) a_x
\ee
with the two different hopping parameters $v_1$ and $v_2$.

We consider the case of constant magnetic field originated from  potential
$$
A_1(x_1,x_2) = B x_2, \quad A_2(x_1,x_2) = 0
$$
The values of magnetic flux through the lattice cell are chosen to be equal to $1/N$ of elementary flux $\Phi_0 = h/e$. The two possibilities are considered: with $N = 3$ and $N=4$. We calculate numerically spectrum of the system, and the value of Hall conductivity using the obtained above expression
\be
\bar \si^{AS}_{ij}&=
\epsilon_{ij}\, \frac{e^2}{h} \, \frac{1}{N}\, \frac{\epsilon^{abc}}{24  \pi^2} \,
\int d\om \intl_{BZ} dp_1  dp_2 \Tr
\[
\frac{\pd {\bm Q}_{p_1p_2}^\om}{\pd p_a}
{\bm G}_{p_1p_2}^\om
\frac{\pd {\bm Q}_{p_1p_2}^\om}{\pd p_b}
{\bm G}_{p_1p_2}^\om
\frac{\pd {\bm Q}_{p_1p_2}^\om}{\pd p_c}
{\bm G}_{p_1p_2}^\om
\]\label{sHall}
\ee

In both cases we reproduce the known result for spectrum (that results from the solution of Harper equation). Moreover, in the case when the chemical potential belongs to the gap, we reproduce the value of Hall conductivity that might me obtained alternatively using the solution of Diophantine equation. This confirmes the validity of the derived general expression for the Hall conductivity as an integral in the Brillouin zone.

Recall that the standard method for calculation of Hall conductivity gives
$$\sigma_H = \frac{e^2}{h}\,t_r$$
 where $t_r$ is the solution of Diophantine equation
 $$
 r = N s_r + \nu t_r
 $$
for integer $r,s_r,t_r$. In our case $N = 3$ or $4$ while $\nu = 1$. Here $|t_r| \le N/2$, while $r = 1,..., N$.

One can check that both for $N=3$ and $N=4$ there are
two nontrivial solutions of this equation with $t_r = \pm 1$. This pattern is reproduced by our numerical results and is illustrated by Fig. \ref{fig.topoNum_3} and Fig. \ref{fig.topoNum_4}.

\subsection{The case of N=3}

In this section, we consider the topological number $\sigma_H$ defined above, in the
presence of magnetic field $\cal B$, which satisfies ${\cal B}a^2=\phi_0/3$.
Here, $\phi_0$ is the magnetic flux quantum,
$\phi_0=h/e$. For simplicity we use here unities with $\hbar = e = 1$. In these units the value of conductivity is quantized as an integer multiple of $1/(2\pi)$.

If one switches on a small electric field $\cal E$ along $x$-axis, according to Eq. (\ref{sHall})
the Hall current density along $y$-axis $J_2$ is given by
\begin {eqnarray}\label{Hall_current_density}
J_2=\int_{-\infty}^{+\infty}\frac{dE}{2\pi}
\int_{-\pi/3}^{+\pi/3}\frac{d p_1}{2\pi}
\int_{-\pi}^{+\pi}\frac{d p_2}{2\pi}
G(p) \frac{\partial Q}{\partial p_2}
\end{eqnarray}
where $Q$ is the matrix
\begin {eqnarray}\label{Q_matrix}
Q= \begin{pmatrix}
E - 2 v_1 cos(p_1)  & -v_2 e^{ip_2}         & -v_2 e^{-ip_2}      \\
-v_2 e^{-ip_2}   & E - 2 v_1 cos(p_1+2\pi/3) &  -v_2 e^{ip_2}     \\
-v_2 e^{ip_2}    & -v_2 e^{-ip_2}    & E -2 v_1 cos(p_1-2\pi/3)
\end{pmatrix}
\end{eqnarray}
In the leading order, $G=G^{(1)}(p)$ which is given by
\begin {eqnarray}\label{G1}
G^{(1)}(p)=\frac{i}{2} G^{(0)}\frac{\partial Q}{\partial p_i}
\frac{\partial G^{(0)}}{\partial p_j} F_{ij},
\end{eqnarray}
with $ G^{(0)}=1/Q$ and $i,j\in \{0,1\}$. Therefore,
$G^{(1)}=(i/2) G^{(0)}(\partial_1 Q \partial_0 G^{(0)}
-\partial_0 Q \partial_1 G^{(0)})(i \cal E)$, and
\begin {eqnarray}\label{Hall_current_density_2}
J_2=\frac{-{\cal E}}{2}
\int_{-\infty}^{+\infty}\frac{dE}{2\pi}
\int_{-\pi/3}^{+\pi/3}\frac{d p_1}{2\pi}
\int_{-\pi}^{+\pi}\frac{d p_2}{2\pi}
Tr G^{(0)}(\frac{\partial Q}{\partial p_1}\frac{\partial G^{(0)}}{\partial p_0}
-\frac{\partial Q}{\partial p_0}\frac{\partial G^{(0)}}{\partial p_1})
\frac{\partial Q}{\partial p_2}.
\end{eqnarray}

With the help of Mathematica, we obtained
\begin {eqnarray}\label{trace}
&& {\rm Tr} \Big( G^{(0)}\big(\frac{\partial Q}{\partial p_1}\frac{\partial G^{(0)}}{\partial p_0}-
\frac{\partial Q}{\partial p_0}\frac{\partial G^{(0)}}{\partial p_1}\big)
\frac{\partial Q}{\partial p_2} \Big)  \nonumber\\
&=& 6\sqrt{3} u^2 i \, \frac{E-U}{E^3-\frac{12+3u^3}{4}E -V},
\end{eqnarray}
where $U=\frac{u}{4}cos(3p_1)+\frac{1}{2}cos(3p_2)$
and $V=\frac{u^3}{4}cos(3p_1)+2 cos(3p_2)$, with $u=2v_1/v_2$.
Note that if one wants to take chemical potential into account, one will
change $E$ into $E+\mu$. From the denominator of Eq.(\ref{trace}), we
can find the energy spectra of the system. The energy levels will be
the solutions of the cubic equation
$4(E/2\alpha)^3 -3(E/2\alpha)=V/2\alpha^3$,
with $\alpha=\sqrt{u^2+4}/2$.
Explicitly, the solutions (from small to big) are given by
\begin {eqnarray}\label{roots}
E_1&=& 2\alpha \, cos(\frac{\theta+2\pi}{3})  \nonumber\\
E_2&=& 2\alpha \, cos(\frac{\theta-2\pi}{3})  \nonumber\\
E_3&=& 2\alpha \, cos(\frac{\theta}{3}),
\end{eqnarray}
with $\theta= arccos(V/2\alpha^3)\in [0,\pi]$.

\begin{figure}[h]
\centering  %
\includegraphics[width=10cm]{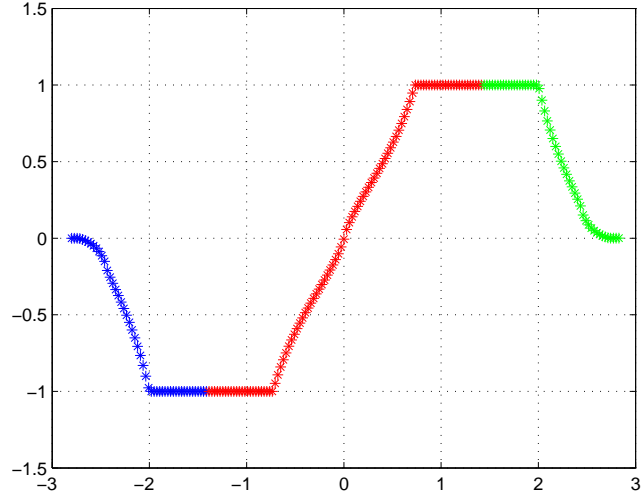} %\vspace{1cm} %
\caption{Topological number v.s. chemical potential, for the case of $N=3$ and $u=2$.}  %
\label{fig.topoNum_3}   %
\end{figure}

Next step is to calculate the Hall conductivity from
Eq.(\ref{Hall_current_density}). Our method is to integrate
$dE$ analytically and then compute the integral $d^2 {\bf p}$
numerically. Taking the case of $\mu< inf\{E_2\} $,
as an example, we replace $E$ by $E+\mu+i\delta_{\bf p}$
with $\delta_{\bf p}=\eta\, \Theta(E_1(p)-\mu)$.
Applying residue theorem to the integration of dE,
we obtained
\begin {eqnarray}\label{Hall_current_density_3}
J_2={\sqrt{3}\cal E}
\int_{-\pi}^{+\pi}\frac{d a}{2\pi}
\int_{-\pi}^{+\pi}\frac{d b}{2\pi}
(E_1-E_2)^{-2}(E_3-E_1)^{-2}[1-2(U-E_1)(\frac{1}{E_1-E_2}-\frac{1}{E_3-E_1})],
\end{eqnarray}
where $E_i$'s are the roots, but as the functions of variables $a=3p_1$ and $b=3p_2$.
Then using Matlab, we obtained the numerical results shown in Fig.\ref{fig.topoNum_3},
via numerical integration.

% spectra_4

%
\begin{figure}[h]
\centering  %
\includegraphics[width=10cm]{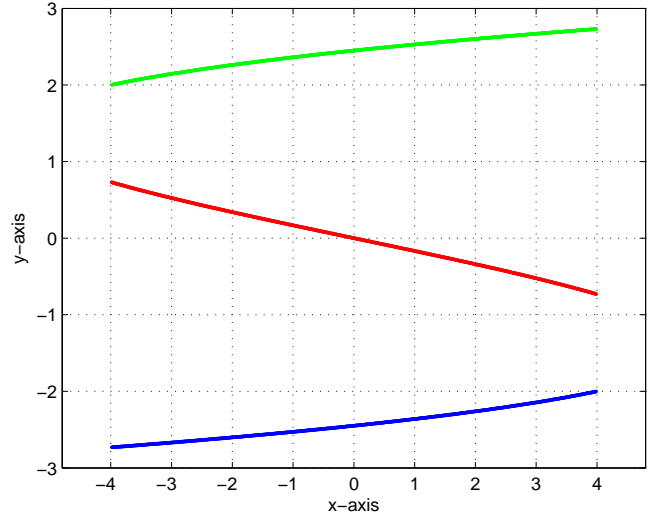} %\vspace{1cm} %
\caption{Energy spectra of N=3, with $u=2$. In the x-axis: $x=V$.}  %
\label{fig.spectra_3}   %
\end{figure}

\subsection{The case of N=4}

In this section, we consider the case of $N=4$, i.e.
the magnetic field $\cal B$ satisfies ${\cal B}a^2=\phi_0/4$,
with $\phi_0=h/e$.
If one switches on a small electric field $\cal E$ along $x$-axis,
the Hall current density along $y$-axis $J_2$ is given by
\begin {eqnarray}\label{Hall_current_density_q=4}
J_2=\int_{-\infty}^{+\infty}\frac{dE}{2\pi}
\int_{-\pi/4}^{+\pi/4}\frac{d p_1}{2\pi}
\int_{-\pi}^{+\pi}\frac{d p_2}{2\pi}
G(p) \frac{\partial Q}{\partial p_2}
\end{eqnarray}
where $Q$ is the matrix
\begin {eqnarray}\label{Q_matrix}
Q= \begin{pmatrix}
E - 2 v_1 cos(p_1)  & -v_2 e^{ip_2}           &0                      & -v_2 e^{-ip_2}      \\
-v_2 e^{-ip_2}      & E - 2 v_1 cos(p_1+\pi/2) &  -v_2 e^{ip_2}       & 0        \\
0                   & -v_2 e^{-ip_2}         & E -2 v_1 cos(p_1+\pi)  &   -v_2 e^{ip_2} \\
-v_2 e^{ip_2}      & 0                      & -v_2 e^{-ip_2}          & E -2 v_1 cos(p_1-\pi/2)
\end{pmatrix}
\end{eqnarray}

Using Mathematica, we obtained
\begin {eqnarray}\label{trace_q=4}
&& {\rm Tr} \Big( G^{(0)}\big(\frac{\partial Q}{\partial p_1}\frac{\partial G^{(0)}}{\partial p_0}-
\frac{\partial Q}{\partial p_0}\frac{\partial G^{(0)}}{\partial p_1}\big)
\frac{\partial Q}{\partial p_2} \Big)  \nonumber\\
&=& 16 u^2 i \, \frac{E(E^2+U)}{[E^4-(u^2+4)E^2 + V]^2},
\end{eqnarray}
where $U=\frac{u^2}{4}(1-cos(4p_1))+(1-cos(4p_2))$
and $V=\frac{u^4}{8}(1-cos(4p_1))+2 (1-cos(4p_2))$, with $u=2v_1/v_2$.
Note that if one wants to take chemical potential into account, one will
change $E$ into $E+\mu$. From the denominator of Eq.(\ref{trace_q=4}), we
can find the energy spectra of the system. The energy levels will be
the solutions of the quadratic equation
$E^4 -(u^2+4)E^2+V=0$.
Explicitly, the solutions (from small to big) are given by
\begin {eqnarray}\label{roots_q=4}
E_1&=& -(w+\sqrt{w^2-V})^{1/2}  \nonumber\\
E_2&=& -(w-\sqrt{w^2-V})^{1/2}  \nonumber\\
E_3&=& (w-\sqrt{w^2-V})^{1/2}  \nonumber\\
E_4&=& (w+\sqrt{w^2-V})^{1/2}  \nonumber\\
\end{eqnarray}
with $w=u^2/2+2$.

\begin{figure}[h]
\centering  %
\includegraphics[width=10cm]{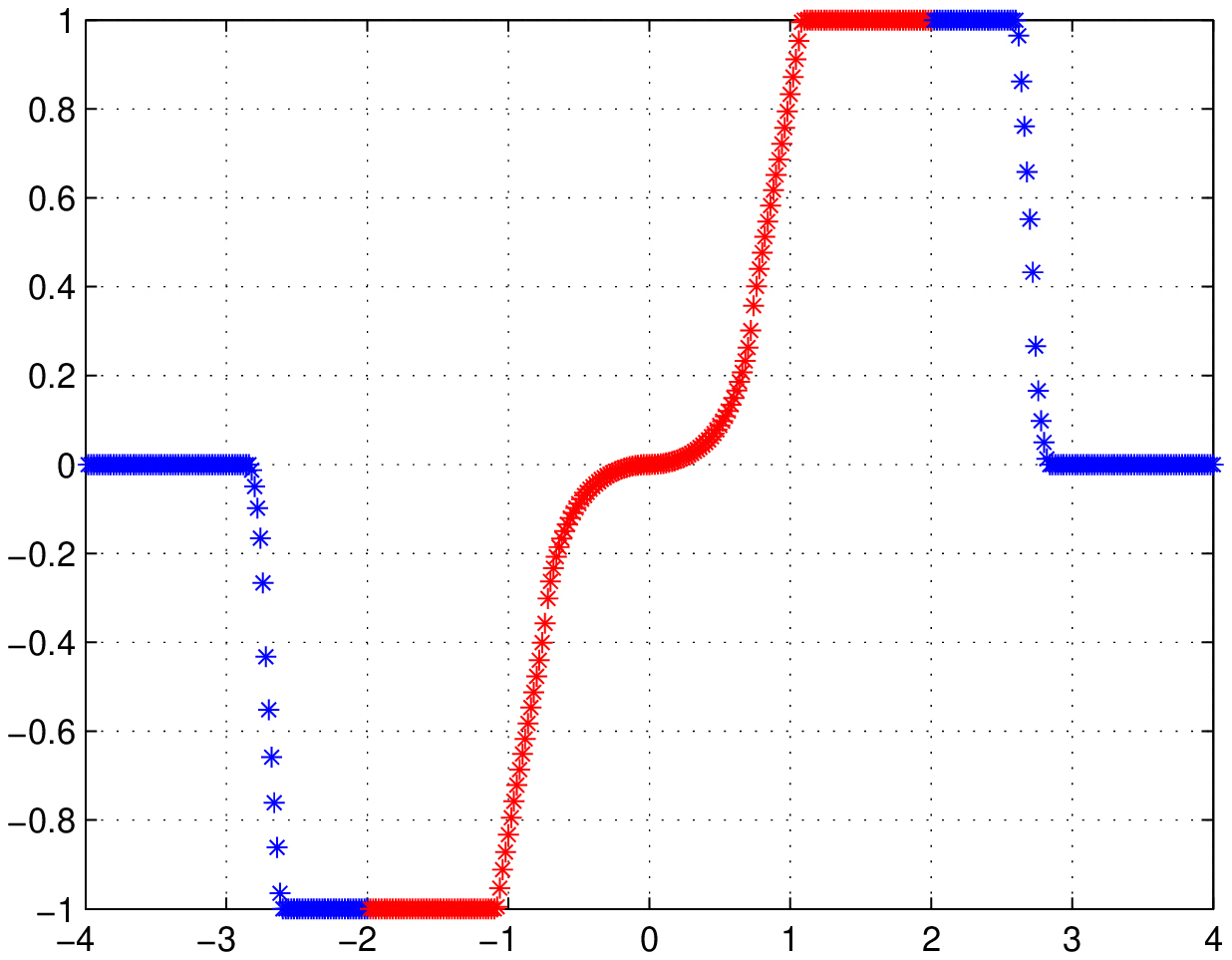} %\vspace{1cm} %
\caption{Topological number v.s. chemical potential, for the case of $N=4$ and $u=2$.}  %
\label{fig.topoNum_4}   %
\end{figure}

Next step is to calculate the Hall conductivity from
Eq.(\ref{Hall_current_density_q=4}). Our method is to integrate
$dE$ analytically and then compute the integral $d^2 {\bf p}$
numerically. Taking the case of $\mu< inf\{E_2\} $,
as an example, we replace $E$ by $E+\mu+i\delta_{\bf p}$
with $\delta_{\bf p}=\eta\, \Theta(E_1(p)-\mu)$.
Applying residue theorem to the integration of dE,
we obtained
\begin {eqnarray}\label{Hall_current_density_q=4_b}
J_2={2u^2\cal E}
\int_{-\pi}^{+\pi}\frac{d a}{2\pi}
\int_{-\pi}^{+\pi}\frac{d b}{2\pi}
(E_1-E_2)^{-2}(E_1-E_3)^{-2}(E_1-E_4)^{-2}
[3E_1^2+U-2(E_1^3+UE_1)H],
\end{eqnarray}
where $H=\frac{1}{E_1-E_2}+\frac{1}{E_1-E_3}+\frac{1}{E_1-E_4}$,
and $E_i$'s are the roots, but as the functions of variables $a=4p_1$ and $b=4p_2$.
Then using Matlab, we obtained the numerical results shown in Fig.\ref{fig.topoNum_4} ,
via numerical integration.

% spectra_4

%
\begin{figure}[h]
\centering  %
\includegraphics[width=10cm]{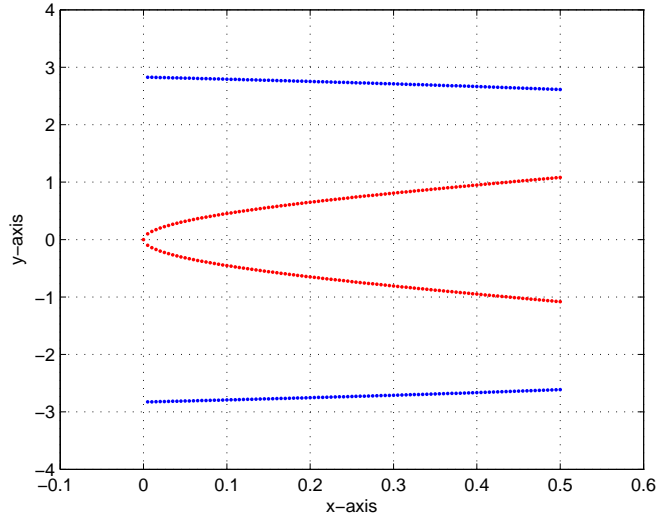} %\vspace{1cm} %
\caption{Energy spectra of N=4, with $x=V/w^2$, and $u=2$.}  %
\label{fig.spectra_4}   %
\end{figure}

\section{Conclusions and discussion}
\label{SectionV}

In the present paper we considered the non - interacting fermionic systems in two space dimensions. We restrict our consideration to the rectangular lattices. The consideration is illustrated by the model with the simplest one - particle Hamiltonian of Eq. (\ref{TBH}). However, our results are valid for the tight - binding Hamiltonian of a more general type.

We consider the given systems in the presence of constant external magnetic field such that the magnetic flux through the lattice cell is equal to
$$
\Phi = \frac{\nu}{N} \times \Phi_0
$$
where $\Phi_0$ is a quantum of magnetic flux $\Phi_0 = h/e$.

We divide the Brillouin zone into the Magnetic Brillouin zones with eigenvectors of momentum
 \be
\ket{un,v}\equiv  \Ket{u +n\frac{2\pi}{aN}\nu,v}\label{bv}
\ee
Here $u \in [\Delta, \Delta + \frac{2 \pi }{a N})$ is momentum along the $x_1$ axis while $v$ is momentum along the $x_2$ axis, while $n = 0,..., N-1$. $\Delta$ may be arbitrary. Harper representation of Hamiltonian is its representation in the basis of Eq. (\ref{bv}). In this representation Hamiltonian $\hat{H}$ is diagonal in $u$ and $v$ but is not diagonal in $n$. For the simplest tight - binding model of Eq. (\ref{TBH}) the lattice Dirac operator ${\bf Q} = -i\om + \hat{H}$ in Harper representation has nonzero matrix elements given by Eq. (\ref{QHarper}) with the transitions between the adjacent values of $n$. In a more general case the lattice Dirac operator in Harper representation has the form with the transitions between any possible pairs of $n$, but remains diagonal in $u$ and $v$. Dirac operator $\bf Q$ becomes the $N\times N$ matrix depending on $u$ and $v$. The Green function ${\bf G}= {\bf Q}^{-1}$ also becomes the $N\times N$ matrix. Both $\bf Q$ and $\bf G$ defined in this way belong to the magnetic Brillouin zone,  which is $N$ times smaller than the whole Brillouin zone, i.e. $u \in [\Delta, \Delta + \frac{2 \pi }{a N})$, while $v \in [0, \frac{2 \pi }{a })$. It is worth mentioning that matrix $\bf Q$ obeys specific boundary conditions that depend on $n$.

However, we are able to extend the definition of matrices $\bf Q$ and $\bf G$ to the whole Brillouin zone. After this extension matrix $\bf Q$ obeys periodic boundary conditions in the whole Brillouin zone. As a result we are able to represent the Hall conductivity of the given system as follows
 $$
 \boxed{\sigma_H = \frac{e^2}{h} \, \frac{1}{N}\,{\cal N}}
 $$
Here $\cal N$ is the topological invariant composed of the $N\times N$ matrices $\bf Q$ and $\bf G$:
\be
\boxed{
	{\cal N} =
	\frac{\epsilon^{abc}}{24  \pi^2} \,
	\int d\om \intl_{BZ} dp_1  dp_2 \Tr
	\Bigl[
	\frac{\pd {\bm Q}_{p_1p_2}^\om}{\pd p_a}
	{\bm G}_{p_1p_2}^\om
	\frac{\pd {\bm Q}_{p_1p_2}^\om}{\pd p_b}
	{\bm G}_{p_1p_2}^\om
	\frac{\pd {\bm Q}_{p_1p_2}^\om}{\pd p_c}
	{\bm G}_{p_1p_2}^\om
	\Bigr]} \label{NT}
\ee

{\bf This representation is the main result of our paper.} Notice that it is valid for the non - interacting tight - binding fermionic system of general form  defined on rectangular lattice. For the non - interacting systems this invariant {\it should} be equal to the integer multiple of $N$ in order to provide the integer QHE. We check this numerically for the cases with $\nu = 1$, $N = 3,4$. The values of Hall conductivity obtained using solution of Diophantine equation are reproduced as it should be.

The experience of the previously known expressions for the Hall conductivity indicates that Eq. (\ref{NT}) remains valid also for the case of an interacting system. Then matrix $\bf G$ is likely to be replaced by the complete electron propagator in Harper representation (defined originally in the magnetic Brillouin zone, but extended analytically to the whole Brillouin zone).
In this case the value of $\cal N$ is already not necessarily equal to the integer multiple of $N$. This way we come to the fractional value of conductivity that is expressed through the topological invariant.

Thus we propose the hypothesis that the  topological description of fractional quantum Hall effect may be achieved using the above mentioned expression with the electron Green function written in Harper representation. The detailed check of this hypothesis remains the important challenge of the future work and is out of the scope of the present paper.

M.A.Z. is grateful to G.Volovik for useful discussions, which became the starting point of the present work. M.S. is grateful to I.Fialkovsky for numerous useful private communications.

\appendix

\section{Coherent states for bosonic systems }

\label{AppendixA}
\subsection{Bosons - single particle: Hilbert space}

In this section for completeness we introduce notations to be used widely in the remaining text. We start from the standard definitions of the coherent states.

Dealing with the harmonic oscillator, i.e. with the system containing the tower of the one - particle states, we define the creation operator:
%\subsubsection*{Coherent states - bosons}
\be
a^\+\ket{n}=\sqrt{n+1}\ket{n+1} \tab
a^\+\ket{0}=\sqrt{1}\ket{1} \tab a^\+\ket{1}=\sqrt{2}\ket{2}
\ee
and correspondingly
\be
\ket{2}=\frac{a^\+ a^\+}{\sqrt{2!}}\ket{0}
\ee
and
\be
\ket{n}=\frac{(a^\+)^n}{\sqrt{n!}}\ket{0}
\ee
The annihilation operator acts as
\be
a\ket{n}=\sqrt{n}\ket{n-1}
\ee
The coherent state is defined as the state $\ket{\al}$ such that
\be
a\ket{\al}=\al\ket{\al}
\ee
One can represent
\be
\ket{\al}=\sum_{n=0}^\infty c_n \ket{n}
\ee
Therefore,
\be
a\ket{\al}=\sum_{1} c_n \sqrt{n}\ket{n-1}=
\sum_{0} c_{n+1} \sqrt{n+1}\ket{n}
\ee
and
\be
c_{n+1}=\frac{\al c_n}{\sqrt{n+1}}
\ee
In particular,
\be
c_1=\frac{\al c_0}{\sqrt{1}} \tab
c_2=\frac{\al c_1}{\sqrt{2}} = \frac{\al^2 c_0}{\sqrt{2!}} \tab...\tab
c_n=\frac{\al^n c_0}{\sqrt{(n+1)!}}
\ee
We come to expression
\be
\ket{\al}=\sum_{n=0}^\infty \frac{\al^n c_0}{\sqrt{(n+1)!}} \ket{n}
=c_0 \sum_{n=0}^\infty  \frac{(\al a^\+)^n }{(n+1)!} \ket{0}=
c_0 e^{\al a^\+} \ket{0}
\ee
and
\be
\bra{\al}=\bra{0} c_0^* e^{\al^* a}
\ee
Thererore,
\be
\braket{\al|\beta} = |c_0|^2 \bra{0} c_0^* e^{\al^* a} \ket{\beta}=
|c_0|^2 \bra{0} c_0^* e^{\al^* \beta} \ket{\beta}=
|c_0|^2 e^{\al^* \beta}
\ee
we can set $c_0=1$
\be
\ket{\al}=e^{\al a^\+}\ket{0} \tab
\bra{\al}=\bra{0} e^{\al^* a} \tab
\braket{\al|\beta}=e^{\al^*\beta}
\ee

\subsection{Bosons: Fock space}

Let us consider many - particle bosonic states.
Single state creation operator acts on vacuum as follows
\be
\ket{0,...,1_\la,...,0}\equiv\ket{\lambda} = a^\dag_\lambda\ket{0} \tab
\bra{\la}=\bra{0} a_\lambda
\ee
and
\zz{\be
	\ket{0,...,n_\la,...,0}\equiv\ket{\lambda^n} = \frac{(a_\lam^\+)^n}{\sqrt{n!}}\ket{0}
	\ee
	In addition we define
	\be
	\ket{n_1,...,n_\la,...}\equiv\ket{n_\lambda} = \frac{(a_1^\+)^{n_1}}{\sqrt{n_1!}}...\frac{(a_\lam^\+)^{n_\la}}{\sqrt{n_{\la}!}}...\ket{0}
	\ee }

According to the completeness relation we have
\zz{\be
	\sum_{n_{\lam}} \ket{{n_{\lam}}}\bra{{n_{\lam}}}=1
	\ee }
Fock space coherent state is defined as
\be
\ket{\psi}=\exp\({\sum_i \psi_i a_i^\dag}\)\ket{0}
\ee
where $a_i^\dag$ creates the "$i$"state
\be
a_i\ket{\psi}=\psi_i\ket{\psi} \tab
\bra{\psi}a_i^\dag=\bra{\psi}\bar\psi_i
\ee
\be
a_i^\dag\ket{\psi}=\pd_{\psi_i}\ket{\psi}
\tab
\bra{\psi}a_i=\pd_{\bar\psi_i}\bra{\psi}
\ee
The inner product of two coherent states is given by
\be
\braket{\psi'|\psi}=\exp\(\sum_i\bar\psi'_i\psi_i\)
\ee
Projection of the coherent state to the one - particle state is
\be
\braket{\al|\psi}=\bra{0}a_\al\ket{\psi}=\psi_\al \tab
\braket{\psi|\al}=\bra{\psi}a_\al^\+\ket{0}=\bar\psi_\al
\ee
The completeness relation may be represented as
\be
\int d\(\bar\psi,\psi\) e^{-\sum_i\bar\psi_i\psi_i}\ket{\psi}\bra{\psi}=1
\ee
where
\be
d\(\bar\psi,\psi\)\equiv
\prod_i \frac{d\bar\psi_id\psi_i}{\pi}
\ee
\subsection{Changing basis for one - particle creation and annihilation operators}

%\be
%a_\al^\+\ket{0}=\ket{\al}=\sum_{n_\beta} %\ket{n_\beta}\braket{n_\beta|\al}=
%\sum_\beta a^\+_\beta\ket{0} %\braket{\beta|\al}
%\ee

%\be
%a_\al^\+=
%\sum_\beta a^\+_\beta \braket{\beta|\al}
%\ee
%\be
%\bra{0} a_\al =\bra{\al}=
%\sum_\beta \braket{\al|\beta} \bra{\beta}=
%\sum_\beta \braket{\al|\beta} %\bra{0}a_\beta
%\ee
%\be
%a_\al=\sum_\beta \braket{\al|\beta} a_\beta
%\ee
%\be
%\sum_\al a_\al^\+ a_\al =
%\sum_\al
%\sum_\la \braket{\la|\al} a_\la^\+
%\sum_{\la'} \braket{\al|\la'} a_{\la'}=
%\sum_{\la} \sum_{\la'} \braket{\la|\la'} %a_\la^\+ a_{\la'}
%\ee
Let us consider the 1D system defined in the interval $(0,L)$, where $\braket{x|k}=\braket{k|x}^*=\frac{e^{ikx}}{\sqrt{L}}$
We define the creation and annihilation operators in momentum space
\be
a^\+(k)=\int_0^Ldx a^\+(x) \braket{x|k}=
\int_0^L dx a^\+(x)\frac{e^{ikx}}{\sqrt{L}}
\ee
\be
a(k)=\int_0^Ldx\braket{k|x}a(x)=
\int_0^Ldx\frac{e^{-ikx}}{\sqrt{L}}a(x)
\ee
The inverse transformation gives
\be
a^\+(x)=\sum_k a^\+(k) \braket{k|x}=
\sum_k a^\+(k) \frac{e^{-ikx}}{\sqrt{L}}
\label{a(x)_to_a(k)(dag)}
\ee
\be
a(x)=\sum_k \braket{x|k}a(k)=
\sum_k \frac{e^{ikx}}{\sqrt{L}} a(k)
\label{a(x)_to_a(k)}
\ee
For the system with lattice with spacing $a$ we have instead $\braket{x|k}=\braket{k|x}^*=\frac{e^{ikx}}{\sqrt{2\pi/a}}$
\be
a^\+(x)=\int_0^{2\pi/a}dk a^\+(k) \braket{k|x}=
\int_0^{2\pi/a}dk a^\+(k) \frac{e^{-ikx}}{\sqrt{2\pi/a}}
\ee
\be
a(x)=\int_0^{2\pi/a}dk\braket{x|k}a(k)=
\int_0^{2\pi/a}dk\frac{e^{ikx}}{\sqrt{2\pi/a}}a(k)
\ee
\be
a^\+(k)=\sum_x a^\+(x) \braket{x|k}=
\sum_x a^\+(x) \frac{e^{ikx}}{\sqrt{2\pi/a}}
\ee
\be
a(k)=\sum_x \braket{k|x}a(x)=
\sum_x \frac{e^{-ikx}}{\sqrt{2\pi/a}} a(x)
\ee
\be
\sum_x a_x^\+ a_x=
\int_0^{2\pi/a} dk \int_0^{2\pi/a} dk'
%\frac{1}{2\pi/a}
\braket{k|k'} a_k^\+ a_{k'}=
% \frac{1}{2\pi/a}
\int_0^{2\pi/a} dk \,\, a_k^\+ a_{k}
\ee
\subsection{Changing basis for the  Coherent states}
\be
\ket{\psi}=\exp\({\sum_\al \psi_\al a_\al^\dag}\)\ket{0}=
\exp\({\sum_\al \psi_\al \sum_\beta \braket{\beta|\al} a^\+_\beta}\)\ket{0}=
\exp\(\sum_\beta \psi_\beta a_\beta^\+ \)
\ee
where we define
\be
\psi_\beta \equiv \sum_\al \psi_\al\braket{\beta|\al}
\ee
That means there is no need to label the coherent state
\be
\boxed{
	\ket{\psi}=\ket{\psi_\al}=\ket{\psi_\beta}=\ket{\psi_x}=\ket{\psi_k}
}
\ee

%\be
%\ket{\psi}=\sum_\la \ket{\la}\braket{\la|\psi}=
%\sum_\la \ket{\la}\bra{0}a_\la\ket{\psi}=
%\sum_\la \ket{\la}\psi_\la
%\ee
For the space Fourier transforms on the lattice
\be
\psi_x=\int_0^{2\pi/a}dk\braket{k|x}\psi_k=
\int_0^{2\pi/a}dk\frac{e^{-ikx}}{\sqrt{2\pi/a}}\psi_k
\ee
\be
\bar\psi_x=\int_0^{2\pi/a}dk\braket{x|k}\bar\psi_k=
\int_0^{2\pi/a}dk\frac{e^{+ikx}}{\sqrt{2\pi/a}}\bar\psi_k
\ee
\be
\psi_k=\sum_x \braket{x|k} \psi_x=
\sum_x \frac{e^{ikx}}{\sqrt{2\pi/a}} \psi_x
\ee
\be
\bar\psi_k=\sum_x \braket{k|x} \bar\psi_x=
\sum_x \frac{e^{-ikx}}{\sqrt{2\pi/a}} a\bar\psi_x
\ee

\section{Coherent states for fermionic systems}
\label{AppendixB}
\subsection{Single particle fermion states: Hilbert space}
Creation and annihilation operators obey anti - commutation relations
\be
\{a^+,a\}=a^\+a+aa^\+=1
\ee
Action of these operators is given by
\be
a^\+\ket{0} =\ket{1} \tab a^\+\ket{1} =0\\
a\ket{0} =0 \tab a\ket{1} = \ket{0}
\ee
Grassman variables obey the commutation relations
\be
\{C_i,C_j\}=C_iC_j+C_jC_i=0 \tab C_i^2=0
\ee
Arbitrary function of the Grassman variables may be written as
\be
f(C_1,C_2)=a_0+a_1C_1+a_2C_2+a_3C_1C_2
\ee
where $a_0,...,a_3$ are ordinary c-numbers\\
Left differentiation is defined as
\be
\frac{\pd^L f}{\pd C_1}=a_1+a_3C_2
\tab
\frac{\pd^L f}{\pd C_2}=a_2-a_3C_1
\ee
while right differentiation is
\be
\frac{\pd^R f}{\pd C_1}=a_1-a_3C_2
\tab
\frac{\pd^R f}{\pd C_2}=a_2+a_3C_1
\ee
Integration is defined as
\be
\(\int dC\)^2= \int dC_1 \int dC_2 =\int dC_2 \int dC_1=-\int dC_1 \int dC_2
\ee
hence
\be
\int dC=0
\ee
we may choose
\be
\int dC C=1
\ee
Therefore,
\be
\int dC_1 f(C_1,C_2) = a_1 + a_3C_2 \tab
\int dC_2 f(C_1,C_2) = a_1 - a_3C_1
\ee

\be
\int dC_1 \int dC_2 f(C_1,C_2) = - a_3 \tab
\int dC_2 \int dC_1 f(C_1,C_2) =  a_3
\ee
Coherent state is to be constructed as follows
\be
a\ket{\eta}=\eta\ket{\eta}
\ee
where $\eta$ is a Grassman variable. We may postulate
\be
\{\eta,a\}=\{\eta,a^\+\}=0
\ee
hence
\be
\eta\ket{0}=\ket{0}\eta \tab  \eta\ket{1}=-\ket{1}\eta
\ee
Fermionic coherent states may be defined as
\be
\ket{\eta}=\ket{0}-\eta \ket{1}=\(1-\eta a^\+\)\ket{0}=
e^{-\eta a^\+}\ket{0}
\ee
Then
\be
a\ket{\eta}=a\(\ket{0}-\eta \ket{1}\) =\eta\ket{0}=\eta\(\ket{0}-\eta\ket{1}\)=\eta\ket{\eta}
\ee
and
\be
\bra{\bar\eta}=\bra{0}-\bra{1}\bar\eta=\bra{0}\(1-a\bar\eta\)=\bra{0} e^{-a\bar\eta}
\ee
\be
\bra{\bar\eta}a^\+=\(\bra{0}-\bra{1}\bar\eta\)a^\+=\bra{0}\bar\eta=
\(\bra{0}-\bra{1}\bar\eta\)\bar\eta=\bra{\bar\eta}\bar\eta
\ee
where $\eta$ and $\bar\eta$ are independent Grassman variables.

In the following we will also use the alternative notations
\be
\bra{\eta}=\(\bra{0}-\bra{1}\bar\eta\)
\ee
in which $\bar \eta$ is replaced by $\eta$ in the bra - vectors.

Scalar product of the two coherent states is
\be
\braket{\eta|\eta}=\(\bra{0}-\bra{1}\bar\eta\) \(\ket{0}-\eta\ket{1}\)=\braket{0|0}+\bar\eta\eta\braket{1|1}=1+\bar\eta\eta=e^{\bar\eta\eta}
\ee

Using the integration rules for Grasmann variables defined above we obtain
\be
\int d\bar\eta d\eta \braket{\bar\eta|\eta}=
\int d\bar\eta d\eta (1+\bar\eta\eta)=-1
\ee
\be
\int d\bar\eta d\eta e^{-\bar\eta A \eta}  =
\int d\bar\eta d\eta \(1-\bar\eta A \eta\) = A
\ee
\be
\bra{\bar\eta}a^\+a\ket{\eta}=\bra{\bar\eta}\bar\eta\eta\ket{\eta}=\bar\eta\eta\braket{\bar\eta|\eta}
=\bar\eta\eta\(1+\bar\eta\eta\)=\bar\eta\eta
\ee
\be
\int d\bar\eta d\eta \bra{\bar\eta} a^\+a\ket{\eta}= \int d\bar\eta d\eta \bar\eta\eta =-1
\ee
\be
\bra{\bar\eta}a^\+a a^\+a\ket{\eta}=
\bra{\bar\eta}\bar\eta a a^\+ \eta\ket{\eta}=
\bra{\bar\eta}\bar\eta \(1- a^\+a\) \eta\ket{\eta}=
\bar\eta\eta\braket{\bar\eta|\eta}-\bra{\bar\eta}\bar\eta^2\eta^2\ket{\eta}=
\bar\eta\eta
\ee
In case of two sets of Grassman variables
\be
\eta=\[\eta_1,...,\eta_N\] \tab \bar\eta=\[\bar\eta_1,...,\bar\eta_N\]
\ee
\be
e^{-\sum_{ij}\bar\eta_i A_{ij}\eta_j}=1+\(-\sum_{ij}\bar\eta_i A_{ij}\eta_j\)+
\frac{1}{2!}\(-\sum_{ij}\bar\eta_i A_{ij}\eta_j\)\(-\sum_{ij}\bar\eta_i A_{ij}\eta_j\)+...
\ee
\be
\int d(\bar\eta,\eta) e^{-\sum_{ij}\bar\eta_i A_{ij}\eta_j}=\det A
\ee
\be
d(\bar\eta,\eta)=\prod_{i=1}^N d\bar\eta_i d\eta_i
\ee
The completeness relation for the fermionic coherent states receives the form
\be
\int d\bar\eta d\eta e^{-\bar\eta\eta} \ket{\eta}\bra{\bar\eta}&=
\int d\bar\eta d\eta \(1-\bar\eta\eta\)
\(\ket{0}-\eta\ket{1}\) \(\bra{0}-\bra{1}\bar\eta\) =\\
&=\int d\bar\eta d\eta \(-\bar\eta\eta\ket{0}\bra{0}+\eta\ket{1}\bra{1}\bar\eta\)=
\ket{0}\bra{0}+\ket{1}\bra{1}=1
\ee
\subsection{Fermions: Fock space }
First of all, let us consider the following  matrix elements
\be
\bra{p}a^\+(k) a(k')\ket{p'}=\bra{0}a(p)a^\+(k) a(k')a^\+(p')\ket{0}
\ee
Using the commutation relations for fermions
\be
\[a_\al,a^\+_\beta\]_+=a_\al a^\+_\beta + a^\+_\beta a_\al=\delta_{\al,\beta}
\ee
we obtain
\be
&\bra{p}a^\+(k) a(k')\ket{p'}=
\bra{0}a(p)a^\+(k) a(k')a^\+(p')\ket{0}=\\
&=\Bra{0}\(\delta(p-k)-a^\+(k) a(p)\) \(\delta(p'-k') a^\+(p') va(k')\)\Ket{0}=\delta(p-k) \delta(p'-k')
\ee
For the coherent state we obtain
\be
\ket{\psi}=\exp\(-{\sum_i \psi_i a_i^\dag}\)\ket{0}=
1\ket{0}-{\sum_i \psi_i a_i^\dag}\ket{0}+\frac{1}{2!}\({\sum_i \psi_i a_i^\dag}\)^2\ket{0}+...
\ee
\be
\bra{\bar\psi}=\bra{0} \exp\(-\sum_i a_i \bar\psi_i\)
\ee

\be
a_i\ket{\psi}=\psi_i\ket{\psi} \tab a_i^\+\ket{\psi}=-\pd_{\psi_i}\ket{\psi}
\ee
\be
\bra{\bar\psi}a^\+_i =\bra{\bar\psi}\bar\psi_i \tab
\bra{\bar\psi}a_i =\pd_{\bar\psi_i} \bra{\bar\psi}
\ee

\be
\braket{\bar\psi'|\psi}&=\bra{0}\exp\(-\sum_i a_i \bar\psi_i'\) \ket{\psi}=\\
&=\bra{0}\(1-\sum_i a_i \bar\psi_i'+\frac{1}{2!}\({\sum_i a_i \bar\psi_i' }\)^2+...
+ \frac{\(-1\)^n}{n!}\({\sum_i a_i \bar\psi_i'}\)^n \)\ket{\psi}
\ee

\be
-a_i \bar\psi_i &= \bar\psi_i a_i\\
+a_i \bar\psi_i a_j \bar\psi_j &= \bar\psi_i \bar\psi_j a_j a_i\\
-a_i \bar\psi_i a_j \bar\psi_j a_k \bar\psi_k &=
\bar\psi_i \bar\psi_j \bar\psi_k a_k a_j a_i
\ee
Hence
\be
\braket{\bar\psi'|\psi}=\exp\(\sum_i \bar\psi_i'\psi_i\)
\ee

\subsection{Normalized coherent states}
\label{AppendixN}
We may define the normalized coherent states denoted by
\be
\kett{\psi}=\exp\(-\frac{1}{2}\sum_i \bar\psi_i\psi_i\)\exp\(-{\sum_i \psi_i a_i^\dag}\)\ket{0}
\ee
\be
\bbra{\psi}=\bra{0} \exp\(-\sum_i a_i \bar\psi_i\) \exp\(-\frac{1}{2}\sum_i \bar\psi_i\psi_i\)
\ee
for which
\be
\bbrakett{\psi}{ \psi}=1
\ee

The integration measure over fermionic fields may be defined containing the exponential factor:
\be
\int d [\bar\psi,\psi ] =
\int \prod_i d\bar\psi_i d\psi_i e^{\sum_i \bar\psi_i\psi_i}= 1
\ee
Action of fermion creation and annihilation operators on the coherent states is:
\be
\bra{\psi} a^\+_i a_j \ket{\psi}= \bra{\psi} \bar\psi_i \psi_j \ket{\psi}=
\bar\psi_i \psi_j e^{\sum_k \bar\psi_k\psi_k}
\ee
The identity - completeness relation may be written in terms of the normalized coherent states
\be
\int d [\bar\psi,\psi ] e^{-\sum_i \bar\psi_i\psi_i} \kett{\psi}\bbra{\bar\psi}=1
\ee
Coordinates and momentum representation on the Lattice:
\be
\braket{x|\psi}=\bra{0}a(x)\ket{\psi}=\psi(x)\braket{0|\psi}=
\psi(x)\bra{0} \exp\({-\sum_x \psi(x) a^\+(x)}\)\ket{0}=\psi(x)
\ee
\be
\braket{k|\psi}=\psi(k)
\ee
\zz{Using completeness relation in Fock space, we obtain:
	\be
	\ket{\psi}=\Big(\ket{0}\bra{0}+ \sum_x \ket{x}\bra{x}+\frac{1}{2!}\sum_{x\ne y} \ket{x,y}\bra{x,y}+... \Big) \ket{\psi}=1+\sum_x \psi(x)\ket{x}+\frac{1}{2!}\sum_{x \ne y} \psi(x)\psi(y) \ket{x,y}+...
	\ee
	and
	\begin{eqnarray}
		\ket{\psi}&=&\Big(\ket{0}\bra{0}+\int_{-\pi/a}^{\pi/a} dk \ket{k}\bra{k}+\frac{1}{2!}\int_{-\pi/a}^{\pi/a} dk_1 dk_2 \ket{k_1,k_2}\bra{k_1,k_2} + ... \Big)\ket{\psi}\nonumber\\ &=&
		1 + \int_{-\pi/a}^{\pi/a} dk \psi(k) \ket{k}+\frac{1}{2!}\int_{-\pi/a}^{\pi/a} dk_1 dk_2\psi(k_1)\psi(k_2) \ket{k_1,k_2} + ...
\end{eqnarray}}

\section{Gaussian integrals}
\label{AppendixC}

\textbf {Real variables}
\be
\int_{-\infty}^{\infty} dx e^{-ax^2}=\(\frac{\pi}{a}\)^{1/2}
\ee
\be
\int_{-\infty}^{\infty} dx e^{-ax^2+bx}=
\int_{-\infty}^{\infty} dx e^{-a\(x-\frac{b}{2a}\)^2-\frac{b^2}{4a}}=
e^{-\frac{b^2}{4a}}\(\frac{\pi}{a}\)^{1/2}
\ee
\be
&\int_{-\infty}^{\infty} d\bm{x} e^{-\bm{x}^T \bm{A x}}
= \int_{-\infty}^{\infty} dx_1...dx_N e^{-x_iA_{ij}x_j}=
\int_{-\infty}^{\infty} dx_1...dx_N
e^{-x_i\(\Omega^{-1} A\Omega\) _{ii}x_i}
=\frac{\(\pi\)^{N/2}}{\(\det \bm{A}\)^{1/2}}
\ee
\be
-\bm{x}^T \bm{A x} + \bm{b}^T\bm{x}=
-\(\bm{x}^T-\frac{1}{2}\bm{b}^T\bm{A}^{-1}\)
\bm{A}
\(\bm{x}-\frac{1}{2}\bm{A}^{-1} \bm{b}\)
+\frac{1}{2}\bm{b}^T\bm{A}^{-1} \bm{b}
\ee
\be
\int_{-\infty}^{\infty} d\bm{x} e^{-\bm{x}^T \bm{A x}+\bm{b}^T\bm{x}}=
e^{\frac{1}{2}\bm{b}^T\bm{A}^{-1} \bm{b}}
\frac{\(\pi\)^{N/2}}{\(\det \bm{A}\)^{1/2}}
\ee
\textbf {Complex variables}
\be
&\int_{-\infty}^{\infty} d\(\bm{z}^\+,\bm{z}\) e^{-\bm{z}^\+ \bm{A z}}
=\frac{\pi^{N}}{\det \bm{A}}
\ee
where $d\(\bm{z}^\+,\bm{z}\)=\prod_i d\Re  z_i d\Im z_i$\\
\be
&\int_{-\infty}^{\infty} d\(\bm{z}^\+,\bm{z}\)
e^{-\bm{z}^\+ \bm{A z} +\bm{b}^\+\bm{z}+\bm{z}^\+\bm{c}}
=e^{\bm{b}^\+\bm{A}^{-1}\bm{c}}\frac{\pi^{N}}{\det \bm{A}}
\ee
since
\be
-\bm{z}^\+ \bm{A z} + \bm{b}^\+\bm{z}+\bm{z}^\+\bm{c}=
-\(\bm{z}^\+ -\bm{b}^\+\bm{A}^{-1}\)
\bm{A}
\(\bm{z}- \bm{A}^{-1} \bm{c}\)
+\bm{b}^\+\bm{A}^{-1} \bm{c}
\ee
\textbf {Grassman variables}
\be
I=\int d(\bm{\bar\eta},\bm{\eta}) e^{-\bm{\bar\eta}^T \bm{A} \bm{\eta}}=\det \bm{A}
\ee
\be
I[\bm{\bar\xi},\bm{\xi}]=\int d(\bm{\bar\eta},\bm{\eta})
e^{-\bm{\bar\eta}^T \bm{A} \bm{\eta}+\bm{\bar\eta\xi} +\bm{\bar\xi\eta}}=
e^{\bm{\bar\xi}^T\bm{A}^{-1} \bm{\xi}} \det \bm{A}
\ee
\be
\frac{1}{I}\frac{\de^2 I[\bm{\bar\xi},\bm{\xi}]}
{\de \bm{\bar\xi} \de\bm{\xi}} \Bigg|_{\bm{\bar\xi}=\bm{\xi}=0}=
\bm{A}^{-1}
\ee

\section{Current operator}
\label{AppendixD}

In order to establish the form of lattice electric current operator we consider the analogy to continuous theory.

The charge density operator in continuous theory is given by
\be
\rho_{charge}(x)=-ea^\+(x)a(x)
\ee
It obeys the continuity equation
\be
\nabla\cdot \^{\vec{J}}=-\pd_t \^\rho
\ee
with the electric current operator $J$.
On the other hand
\be
\pd_t \^\rho = \frac{i}{\hb} \[\^\rho,\^H\]=
-  \frac{ie}{\hb} \[a^\+(x)a(x),\^H\]
\ee

On the lattice we should have the following expression
\be
\nabla \cdot \^{\vec{J}} =\frac{1}{b}\sum_i\(\^J(x)-\^J(x-e_i)\)
\ee
where $b$ is a lattice spacing. Analyzing it we will determine the lattice electric current operator. We have
\be
\sum_i\(\^J_{x}-\^J_{x-e_i}\) =
\frac{ibe}{\hb}  \[a^\+(x)a(x),\^H\]
\ee
The Hamiltonian is
\be
\^H=-t\sum_{j=1,2}\(\^T_j+\^T^\+_j\)
\ee
with
\be
\^T_i=\sum_x a^\+(x) a(x+e_i)
\tab
\^T_i^\+=\sum_x a^\+(x+e_i) a(x)
\ee

We will use commutation relations
\be
&\{a^\+(x),a(y)\}=a^\+(x)a(y) + a(y) a^\+(x) = \delta(x-y)\\
&\{a(x),a(y)\}=\{a^\+(x),a^\+(y)\}=0
\ee
and obtain
\be
&a^\+(x) a(x) a^\+(y) a(z)=\\
&a^\+(x) \(\delta(x-y)- a^\+(y) a(x) \) a(z)=\\
&\delta(x-y) a^\+(x) a(z) - a^\+(x) a^\+(y) a(x) a(z)=\\
&\delta(x-y) a^\+(x) a(z) - a^\+(y) a^\+(x) a(z) a(x) =\\
&\delta(x-y) a^\+(x) a(z) - a^\+(y) \(\delta(x-z)- a(z) a^\+(x) \) a(x) =\\
&\delta(x-y) a^\+(x) a(z) - \delta(x-z) a^\+(y) a(x) +
a^\+(y) a(z) a^\+(x)  a(x)
\ee
hence
\be
\[a^\+(x) a(x) , a^\+(y) a(z)\]=
\delta(x-y) a^\+(x) a(z) - \delta(x-z) a^\+(y) a(x)
\ee
\be
\[a^\+(x) a(x) , T_i\]
&=\sum_y\(\delta(x-y) a^\+(x) a(y+e_i) - \delta(x-(y+e_i)) a^\+(y) a(x)\)=\\
&=a^\+(x) a(x+e_i) - a^\+(x-e_i) a(x)
\ee
\be
\[a^\+(x) a(x) , T_i^\+\]
&=\sum_y\(\delta(x-(y+e_i)) a^\+(x) a(y) - \delta(x-y) a^\+(y+e_i) a(x)\)=\\
&=a^\+(x) a(x-e_i) - a^\+(x+e_i) a(x)
\ee
The commutator with the Hamiltonian is given by
\be
\[a^\+(x) a(x),H\]&=\[a^\+(x)a(x),\sum_i\(T_i+T_i^\+\)\]=\\
&=\sum_i\(a^\+(x) a(x+e_i) - a^\+(x-e_i) a(x) + a^\+(x) a(x-e_i) - a^\+(x+e_i) a(x)\)=\\
&=\sum_i\Bigg(\(a^\+(x) a(x+e_i) -  a^\+(x+e_i) a(x)\) -
\(a^\+(x-e_i) a(x) - a^\+(x) a(x-e_i) \)\Bigg)
\ee

Therefore, we expect that for the properly defined current operator
\be
\sum_i\(\^J_{x}-\^J_{x-e_i}\)=
i\frac{ebt}{\hb}\sum_i\Bigg(\(a^\+(x) a(x+e_i) -  a^\+(x+e_i) a(x)\) -
\(a^\+(x-e_i) a(x) - a^\+(x) a(x-e_i) \)\Bigg)
\ee
We conclude that the current operator may be chosen in the form
\be
J_i(x)=  i\frac{ebt}{\hb} \(a^\+(x) a(x+e_i) -  a^\+(x+e_i) a(x)\)
\ee
In the presence of the electromagnetic field the Hamiltonian is given by
\be
\^H(a^\dag,a)=-t\sum_x\sum_{j=1,2} \(a^\dag_x
e^{-ie b A_j(x)/\hb} a_{x+e_j}+h.c.\) +
\sum_x e a^\+_x A_0(x) a_x
\ee
and the current operator is
\be
\^J_i(x)=  i\frac{ebt}{\hb} \(a^\+(x)  e^{-iebA_i(x)/\hb} a(x+e_i) - a^\+(x+e_i) e^{iebA_i(x)/\hb} a(x)\)
=- \frac{\delta \^H}{\delta A_i(x)}
=- \frac{\delta \^Q}{\delta A_i(x)}
\ee
The electric current summed over the whole lattice is
{
	\be
	\^{\bar J}_i\equiv
	\frac{1}{L^2}\sum_{x} \^J_i(x)=
	-\frac{1}{L^2}\frac{\de \^H}{\de A_i}=
	-\frac{1}{L^2}\frac{\de \^Q}{\de A_i}
	\ee
}
In this case the variation is homogeneous $A_i(x)\ra A_i(x)+\de A_i$. Here, $L$, is the number of the lattice links in the sample, and hence, is dimensionless.

\section{Hall conductivity as a topological invariant }
\label{AppendixE}
Eq. (\ref{sfin}) for electric conductivity may be expressed as
\be
\bar\si_{ij}=- \frac{1}{N} \frac{e^2}{h}\,\frac{1}{4\pi^2}\bm\Tr\pd_i Q G \pd_0 Q G \pd_j Q G
= \bar\si_{ij}^{S} + \bar\si_{ij}^{AS}
\ee
We also express it as
$$
\bar\si_{ij}= -\frac{1}{\gamma} \bm\Tr f_if_0f_j
$$
with $\gamma^{-1} = - \frac{1}{N} \frac{e^2}{h}\,\frac{1}{4\pi^2}$,
where we defined
\be
f_i = (\pd_i Q) G
\ee
Symmetric and antisymmetric (Hall) components of conductivity are given by
\be
-\gamma \bar\si_{ij}^{S}=
\frac{1}{2}\bm\Tr f_if_0f_j + \frac{1}{2}\bm\Tr f_jf_0f_i
\ee
and
\be
-\gamma \bar\si_{ij}^{AS}=
\frac{1}{2}\bm\Tr f_if_0f_j - \frac{1}{2}\bm\Tr f_jf_0f_i
\ee
In these expressions
\be
\bm\Tr f_if_0f_j= \bm\Tr f_0f_jf_i=\bm\Tr f_jf_if_0
\ee
and
\be
\bm\Tr f_if_0f_j=
\frac{1}{3}\bm\Tr f_if_0f_j +
\frac{1}{3}\bm\Tr f_0f_jf_i +
\frac{1}{3}\bm\Tr f_jf_if_0
\ee
Hall conductivity may be expressed as
\be
-\gamma \bar\si_{ij}^{AS}&=
\frac{1}{6}
\Bigg[
\bm\Tr f_if_0f_j+\bm\Tr f_0f_jf_i+\bm\Tr f_jf_if_0-
\bm\Tr f_jf_0f_i-\bm\Tr f_0f_if_j-\bm\Tr f_if_jf_0
\Bigg]=\\
&=\frac{1}{6}
\Bigg[
\(\bm\Tr f_if_0f_j-\bm\Tr f_if_jf_0\)+
\(\bm\Tr f_jf_if_0-\bm\Tr f_0f_if_j\)+
\(\bm\Tr f_0f_jf_i -\bm\Tr f_jf_0f_i\)
\Bigg]=\\
&=\frac{1}{6}
\Bigg[
\de_{ia}\(\de_{0b}\de_{jc}-\de_{0c}\de_{jb}\)+
\de_{ib}\(\de_{0c}\de_{ja}-\de_{0a}\de_{jc}\)+
\de_{ic}\(\de_{0a}\de_{jb}-\de_{0b}\de_{ja}\)
\Bigg]\bm\Tr f_af_bf_c=\\
&=\frac{1}{6} \ep_{i0j}\ep_{abc} \bm\Tr f_af_bf_c
\ee
Variation of this expression is
\be
\de \(\ep_{abc} \bm\Tr f_af_bf_c\)=
\ep_{abc} \bm\Tr \de f_af_bf_c+
\ep_{abc} \bm\Tr f_a \de f_bf_c+
\ep_{abc} \bm\Tr f_af_b \de f_c
\ee
Since
\be
\ep_{abc}\bm\Tr\de f_af_bf_c=
\ep_{abc}\bm\Tr f_c \de f_af_b=
-\ep_{cba}\bm\Tr f_c \de f_af_b=
\ep_{cab}\bm\Tr f_c \de f_af_b=
\ep_{abc}\bm\Tr f_a \de f_bf_c
\ee
we have
\be
\frac{1}{3}\de \(\ep_{abc} \bm\Tr f_af_bf_c\)&=\ep_{abc} \bm\Tr \de f_af_bf_c=
\ep_{abc} \bm\Tr \de \(\pd_a Q G\) \pd_b Q G\pd_c Q G=\\
&=\ep_{abc} \bm\Tr \pd_a \de Q G \pd_b Q G\pd_c Q G+
\ep_{abc} \bm\Tr \pd_a  Q \de G \pd_b Q G\pd_c Q G
\ee
The first term has the form
\be
\ep_{abc} \bm\Tr \pd_a \de Q G \pd_b Q G\pd_c Q G=
-\ep_{abc} \bm\Tr \pd_a \de Q G \pd_b Q \pd_c G
\ee
while
the second term is
\be
&\ep_{abc}\bm\Tr \pd_a  Q \de G \pd_b Q G\pd_c Q G=
\ep_{abc}\bm\Tr \(Q \pd_a G Q\) \(G \de Q G\)  \pd_b Q G\pd_c Q G=\\
&=\ep_{abc}\bm\Tr \pd_a G  \de Q G  \pd_b Q G\pd_c Q=
\ep_{abc}\bm\Tr \de Q G \pd_b Q G\pd_c Q \pd_a G=\\
&=-\ep_{abc}\bm\Tr \de Q \pd_b G\pd_c Q \pd_a G=
-\ep_{abc}\bm\Tr \de Q \pd_a G\pd_b Q \pd_c G
\ee
and
\be
\ep_{abc} \bm\Tr \de \(\pd_a Q G\) \pd_b Q G\pd_c Q G=
-\ep_{abc} \bm\Tr \de \(\pd_a Q G\) \pd_b Q \pd_c G
\ee
We also have
\be
&\ep_{abc} \bm\Tr \de \(\pd_a Q G\) \pd_b Q \pd_c G=
\ep_{abc}\bm\Tr  \(\pd_a\de Q G+\de Q \pd_a G\)\pd_b Q \pd_c G=\\
&=\ep_{abc}\bm\Tr \pd_a \(\de Q  G\)\pd_b Q \pd_c G=
\ep_{abc}\bm\Tr \pd_a \(\de Q  G \pd_b Q \pd_c G\)
-\ep_{abc}\bm\Tr \de Q  G\pd_a \(\pd_b Q \pd_c G\)
\ee
One can check that
\be
\ep_{abc} \pd_a \(\pd_b Q \pd_c G\)=
\ep_{abc} \(\pd_a \pd_b Q\) \pd_c G+
\ep_{abc} \pd_b Q \(\pd_a \pd_c G\)=0
\ee
and using the periodic boundary conditions we obtain
\be
\ep_{abc}\bm\Tr \pd_a \(\de Q  G \pd_b Q \pd_c G\)=0
\ee
Hence the variation of Hall conductivity vanishes
\be
\de \bar\si_{ij}^{AS}&=
\frac{1}{6} \ep_{i0j}\ep_{abc} \de \bm\Tr f_af_bf_c=
\frac{1}{2} \ep_{i0j}\ep_{abc}  \bm\Tr \de f_af_bf_c=0
\ee
This proves that the Hall conductivity is indeed a topological invariant that is not changed under the smooth variation of the system (that leaves it homogeneous).

\end{document}